# CALPHAD modeling and *ab initio* calculations of the Np-U-Zr system

Wei Xie[1, a], Dane Morgan[1, b]

[1]Department of Materials Science and Engineering, University of Wisconsin-Madison, Madison WI 53706, United States

[a] Current address: Department of Materials Science and Engineering, University of California, Berkeley, Berkeley, CA 94720, United States
[b] Corresponding author. Address: 1509 University Avenue, Madison WI 53706, United States. Tel.: + 1 608 265 5879. Fax: +1 608 262 8353. Email address: ddmorgan@wisc.edu.

# Abstract

Phase diagram of Np-U-Zr, a key ternary alloy system of relevance for metallic nuclear fuels, is still largely undetermined. Here a thermodynamic model for the Np-U-Zr system is developed based on Muggianu extrapolation of our models for the U-Zr, Np-Zr, and Np-U binary systems, all employing the CALculation of PHAse Diagrams (CALPHAD) method. This model reproduces available experimental data for the high temperature phases reasonably well, but is uncertain for the lower temperature phases for which experimental data are scarce. *Ab initio* calculations are performed on the BCC phase of the Np-U-Zr system at 28 compositions employing density functional theory (DFT) both in its standard form and the so-called DFT plus Hubbard $U$ (DFT + $U$) modification. When referencing to our CALPHAD model, standard DFT on average overestimates the enthalpy of mixing of the BCC phase by 0.093 eV/atom, while DFT + $U$ reduces the error to 0.009-0.036 eV/atom when the effective Hubbard $U$ parameters are in the ranges of single-structure optimized $U_{eff}$'s determined in our previous studies of the Np-Zr and U-Zr systems (0.65-0.90 for $U_{eff}$(Np) and 0.99-1.49 eV for $U_{eff}$(U), respectively). Our combined CALPHAD and *ab initio* modeling provides an initial attempt to determine the phase diagram of the Np-U-Zr system but further work is needed for a reliable determination.

# 1. Introduction

Actinide rich metallic nuclear fuels are promising candidates for future generation fast nuclear reactors because of their advantages in thermal conductivity, burn-up and recycling over traditional oxides fuels[1]. However, several issues must be addressed in using metallic fuels. One issue is that the melting temperatures of the constituent actinide metals[2]—for example, U (1408 K), Np (917 K), Pu (913 K)—are close to or even lower than many fast reactor design operating temperatures (870-1300 K)[3]. Moreover, metallic fuels have considerable constituent redistribution and interaction with the cladding during burnup. It has been suggested[1] that alloying the actinides with transition metals like Zr and Mo can both boost the melting temperature and suppress the constituent's diffusion and interaction with the cladding. Understanding the underlying mechanisms for such improvements are important for better design and safe and optimal use of the fuels.

Fundamentally, melting temperature, constituent redistribution, and cladding interaction are all strongly tied to the phase stability of the fuels. However, due to their radioactivity, toxicity, and scarcity, experimental measurement of the phase stability of metallic fuels is often quite difficult and expensive. Take U-Pu-Zr-MA (MA=Np, Am, Cm), a promising candidate for metallic fuel as an example. Among all its ternary subsystems, to the best of our knowledge only U-Pu-Zr has systematic experimental studies of their phase diagram available[4]. To gain additional knowledge of these systems, the present authors have previously[5-8] performed *ab initio* calculations on the Np, U, Zr unary and the Np-Zr, U-Zr and Np-U binary systems and developed CALculation of PHAse Diagram (CALPHAD) models for the U-Zr and Np-Zr binary systems. A natural next step is to study their parent ternary system Np-U-Zr, which is reported in this article.

To our knowledge, experimental study of the phase diagram of the Np-U-Zr ternary system has only been reported once in Ref.[9], in which the U-rich corner is studied by Rodríguez, *et al*. around the three temperatures of 793, 868 and 973 K. Due to the limited composition and temperature space explored, the phase diagram of Np-U-Zr is still left largely undetermined. Furthermore, body centered cubic (BCC) Np and Zr are found to be completely miscible in that study[9], which is in disagreement with more recent experimental results obtained on the Np-Zr binary system[10-12]. Rodríguez, *et al.*'s data for Np-Zr[9] have also been found difficult to model consistently with other experimental data[10-12] in our CALPHAD modeling of the Np-Zr binary system[7]. Therefore, in the following our model results will be compared with the experimental data of Rodriguez, *et al*.[9], but the model parameters for Np-U-Zr will not be adjusted to seek better match with the data. Besides Ref.[9], no other experimental thermochemical or phase equilibrium data has been found by us, nor any CALPHAD model for the Np-U-Zr ternary system.

This study has two main purposes. The first is to obtain a CALPHAD model for the Np-U-Zr system. Due to scarcity of experimental data for this ternary actinide system, thermodynamic modeling of the Np-U-Zr ternary system faces substantially more uncertainty than other more common metallic alloy systems, for which experimental data usually abound. To fill the gap, this study aims to develop an initial model that future work can continue to refine when more ternary data become available. Although the current ternary model has considered all phases known to be stable in the binary subsystems, analysis and discussion in this study are mainly focused on the high temperature phases, in particular BCC. The reason is that the high temperature phases have more experimental data for validating our model than for lower temperature ones and are suggested to have weak or no ternary interactions[9], which is consistent with a key assumption of our model. Given the very limited data for validation and absence of ternary interactions in our model, our CALPHAD model is uncertain for the lower-temperature part of the Np-U-Zr system, which is therefore discussed in this article only for the sake of completeness.

The second purpose is to continue to explore the so-called DFT plus Hubbard $U$ (DFT + $U$) approach[13] on this ternary actinide alloy system of Np-U-Zr. Our earlier studies[6, 7] show that calculations using density functional theory (DFT)[14, 15] in its standard form based on the generalized gradient approximation (GGA)[16] tend to overestimate the enthalpies of U-Zr and Np-Zr by about 0.10 and 0.15 eV/atom, and including spin-orbit coupling (SOC) only reduces the error by about 0.02 and 0.03 eV/atom, respectively. These studies[6, 7] also demonstrate that the errors could be reduced using DFT + $U$. Specifically, for U-Zr, the errors are reduced to be about 0.04 eV/atom at $U_{\text{eff}}$ (U) = 1.24 eV (see FIG. 4 of Ref. [6]), while for Np-Zr, the errors are reduced to be about 0.06 eV/atom at $U_{\text{eff}}$(Np) = 0.90 eV (see Fig. 6 of Ref. [7]). Note DFT + $U$ is not applied on Zr so no discussion of $U_{\text{eff}}$(Zr) is necessary. A more recent study[8] of us further finds that for the Np-U binary system, enthalpies from DFT + $U$ in general are also improved or on par with those from standard DFT when $U_{\text{eff}}$ are in the ranges of single-structure optimized $U_{\text{eff}}$'s determined in the two previous studies of the U-Zr[6] and Np-Zr[7] systems (0.65-0.90 for $U_{\text{eff}}$(Np) and 0.99-1.49 eV for $U_{\text{eff}}$(U), respectively; see also section 2.2 of Ref. [8]). The current study therefore seeks to explore to what extent these results can be extended to the Np-U-Zr ternary system, in particular if such $U_{\text{eff}}$'s fitted individually in the binary sub-systems of U-Zr, and Np-Zr can be "transferred" to their ternary parent system. Assessing the extent to which one can transfer the $U_{\text{eff}}$ values among these metallic actinide systems is essential for determining the predictive capability of DFT + $U$ for actinide alloys.

This manuscript proceeds as follows. Section 2 describes the details of the CALPHAD and *ab initio* methodology. Section 3 presents the results and discussion. Finally, Section 4 summarizes this study.

## 2. Methodology

### 2.1. CALPHAD

A thermodynamic model for the Np-U-Zr system is developed by extrapolating the CALPHAD models for the three binary subsystems U-Zr, Np-Zr and Np-U using the Muggianu symmetric scheme[17]. For the U-Zr and Np-Zr binary systems, the models developed and carefully validated in our previous studies[5, 7] are used. For the Np-U system, a model is developed and used based on the experimental data of Mardon and Pearce[18], which to our knowledge is the only experimental study of the phase stability of the Np-U system so far, as reviewed by us before[8]. No new phase is considered in the ternary model beyond those already found in the binary subsystems, which is supported by the finding of no ternary compound in the samples studied by Rodríguez, *et al.*[9]. Moreover, as mentioned in Introduction, all ternary interactions are assumed to be zero (i.e., assuming weak and negligible ternary interactions). For the high temperature BCC and liquid phases, this approximation is justified based on Rodríguez, *et al.*[9]'s observation that the experimentally measured solidus temperatures of Np-U-Zr are "in agreement with the values expected from a linear interpolation of the solidus temperatures of the binary compounds". For the lower temperature phases, however, this extrapolation is uncertain and can potentially introduce significant errors. The resulted ternary model is provided in the Thermo-Calc Database (tdb) format in the Supplementary Material.

The details are exemplified here using the BCC phase. When phase separated, they are labeled γ(Np), γ(U), β(Zr) or their combinations (e.g., γ(Np)+β(Zr) miscibility gap) following the respective conventional labels of the elemental BCC phases. BCC and BCC+BCC' are also used in this work to denote single-phase and two-phase BCC phase fields. The expression for the Gibbs free energy of BCC Np-U-Zr in the Muggianu symmetric scheme[17] is:

$$G_{Np,U,Zr}^{BCC} = (x_{Np}G_{Np}^{BCC} + x_{U}G_{U}^{BCC} + x_{Zr}G_{Zr}^{BCC}) + RT(x_{Np}\ln x_{Np} + x_{U}\ln x_{U} + x_{Zr}\ln x_{Zr})$$
$$+ (x_{Np}x_{U}L_{Np,U}^{BCC} + x_{Np}x_{Zr}L_{Np,Zr}^{BCC} + x_{U}x_{Zr}L_{U,Zr}^{BCC} + x_{Np}x_{U}x_{Zr}L_{Np,U,Zr}^{BCC})$$

(1)

where $x_i$ is the mole fraction, $G_i^{BCC}$ is the Gibbs energy of elemental BCC metal, and $L_{i,j}^{BCC}$ and $L_{i,j,k}^{BCC}$ are the binary and ternary interaction parameter for specie $i/j/k$=Np, U, Zr. Collectively, the first term is the linear Gibbs energy of mixing, the second term the contribution of ideal entropy of mixing to the Gibbs energy, and the third term the excess Gibbs energy of mixing as described by the Redlich–Kister polynomial[19]. The binary interaction parameters $L_{i,j}^{BCC}$ are from the models for the corresponding binary subsystems (Refs.[5, 7] and this work), while the ternary interaction parameters $L_{Np,U,Zr}^{BCC}$ are all zero in the current model. The standard element reference (SER) [20] is used as the Gibbs energy reference state. Finally, note that the CALPHAD parameters in the three binary sub-systems (including the BCC phase) are not from fitting to *ab initio* results but only to experimental data.

## 2.2. *Ab initio*

*Ab initio* calculations are performed on the BCC phase of the Np-U-Zr at 28 compositions, as illustrated in Figure 1. They are modeled using 8 different supercell structures listed in Table 1. Among them, except for the unary structure A for which the one-atom BCC primitive cell is used, all supercells are generated using the special quasi-random structure (SQS) method[21], as implemented in the Alloy Theoretic Automated Toolkit (ATAT)[22, 23]. $A_3B_1$ and $A_1B_1$ binary as well as $A_1B_1C_1$, $A_2B_1C_1$, $A_6B_1C_1$, and $A_2B_3C_3$ ternary structures are generated and tested previously on Mo-Nb, Ta-W and Cr-Fe binary systems in Ref.[24], and on Mo-Nb-Ta and Mo-Nb-V ternary systems in Ref.[25], respectively. Additionally, a $A_4B_3C_1$ ternary structure is generated following the same approach as Ref.[25] and used in this study. For convenience of future reference, the five ternary SQS supercell structures used in this study are provided in the Supplementary Material. The supercell size and the Monkhorst–Pack[26] *k*-point mesh used to sample the Brillouin zone for each supercell are listed in Table 1. For the BCC U-Zr, Np-Zr and Np-U binary supercells, our testing shows that the *k*-point meshes used converge the energy to less than 3 meV/atom, and thus for BCC Np-U-Zr ternary supercells, similar *k*-point meshes adjusted based on supercell sizes to keep an approximately constant density of k-points in reciprocal space are used. Note that the unary and binary structures listed in Table 1 have already been calculated in our previous studies[5-8] and they are included here for completeness.

SQS structures have no short-range order, so their use in this work implicitly makes the assumption that short-range-order has an inessential impact on the thermodynamic models. A rigorous test of this assumption would need cluster expansion, which is computationally expensive and outside the scope of this article. Instead, a check is performed taking BCC U-Zr as testing system by calculating all symmetrically distinct 16-atom cubic supercells at 25, 50 and 75 at.%Zr and comparing them to the corresponding SQS supercells used in this study, which are also 16-atom but monoclinic or triclinic. As found in Figure S1 (Supplementary Material), the SQS are within 0.025 eV/atom of the most stable structures. If assuming that the 16 atom cells explore enough configurational space to capture possible short-range-ordered configurations, then this difference puts an upper bound on the energy stabilization that could be obtained from allowing short-range-order to form. This is because short-range-order is a balance of chemical driven ordering at the cost of reducing entropy. As values from perfectly ordered cells certainly overestimate of the short-range-order impact on the free energy, including the short-range-order should lower energies by significantly less than 0.025 eV/atom, even for the worst case at 50% in the U-Zr system. Therefore, short range order should be a very small effect. This assertion should hold generally true for the whole BCC Np-U-Zr phase based on the similarity of the three binary BCC phases found in our previous studies[5-8].

All *ab initio* calculations are performed in the general framework of spin-polarized DFT[14, 15] using the Vienna *Ab initio* Simulation Package (VASP)[27, 28]. The electron-ion interaction is described with the projector-augmented-wave (PAW) method[29] as implemented by Kresse and Joubert[30]. The PAW potentials used treat $6s^26p^67s^25f^46d^1$, $6s^26p^67s^25f^36d^1$ and $4s^24p^65s^24d^2$ as valence electrons for Np, U and Zr, respectively. The exchange-correlation functional parameterized in the generalized gradient approximation (GGA)[31] by Perdew, Burke and Ernzerhof (PBE)[16] is used. The stopping criteria for self-consistent loops used are 0.1 meV/cell and 1 meV/cell tolerance of total energy for the electronic and ionic relaxation, respectively. Cutoff energy of 450 eV is used throughout all calculations. The partial occupancies are set using the Methfessel-Paxton method[32] of order one with a smearing width of 0.2 eV. The electronic and ionic optimizations are performed using a Davidson-block algorithm[33] and a Conjugate-gradient algorithm[34], respectively.

In our previous studies of the U-Zr and Np-Zr binary systems, spin-orbit coupling (SOC) is found to affect the calculated enthalpy by about 0.02 and 0.03 eV/atom[6, 7], respectively, which are relatively small compared to DFT-GGA's error in enthalpy that are about 0.10 and 0.15 eV/atom, respectively, and hence SOC is not included in this study. Enthalpies values discussed henceforward are by default calculated without SOC unless otherwise noticed.

The DFT+*U* functional proposed by Dudarev *et al.*[13] is used. This form of DFT+*U* does not introduce explicit local exchange *J* term but only an effective Hubbard *U* term that depends on $U_{eff} = U-J$. This approach also recovers the standard DFT functional exactly when $U_{eff} = 0$. DFT + *U* potential is applied only on U and Np sites, but not on any Zr site. For historical reason, *J* is not set to 0 as one conveniently does but instead to 0.6 and 0.51 eV for Np and U, respectively and vary *U* from 0.75 to 2 eV. Putting them together results in a grid of ($U_{eff}$(Np), $U_{eff}$(U)) pairs as illustrated in Figure 2. This detail is provided here merely to explain why results given in this study are at some awkward $U_{eff}$ values like 0.99 and 1.24, which are from $U_{eff}$ = 1.5 - 0.51 and $U_{eff}$ = 1.75 - 0.51. One can reproduce our results as long as the same $U_{eff}$'s are used, regardless of what specific pair of *U* and *J* is used to reach the $U_{eff}$. *Ab initio* data presented next that correspond to ($U_{eff}$(Np), $U_{eff}$(U)) points beyond those marked by filled circle in Figure 2 are generated by spline interpolation and they only serve as guides to the eyes. Note the standard DFT corresponds to the ($U_{eff}$(Np), $U_{eff}$(U)) = (0,0) point on the grid. The metastable solution issue of DFT+*U* is combated using the *U*-ramping method[35] with modifications described in Ref.[6].

Neglecting the pressure dependence, enthalpy (volume) of mixing of BCC Np-U-Zr is defined as:

$$X^{mix}\left(Np_{x_{Np}}U_{x_U}Zr_{x_{Zr}}\right) = X\left(Np_{x_{Np}}U_{x_U}Zr_{x_{Zr}}\right) - x_{Np}X(Np) - x_U X(U) - x_{Zr}X(Zr) \qquad (2)$$

where $X$ is total energy (volume) per unit amount (e.g., per atom), $Np_{x_{Np}}U_{x_U}Zr_{x_{Zr}}$ and Np, U, Zr are the alloy and the constituent BCC elemental metal references, respectively, and $x_{Np}$, $x_U$, $x_{Zr}$ are the mole fractions of Np, U, Zr in the ternary alloy with $x_{Np}+x_U+x_{Zr}=1$.

Following the approach used in Refs. [6, 7] to mitigate the mechanical instability of BCC U-Zr and Np-Zr, only volume relaxation is performed, and the ions are constrained to the ideal BCC lattice sites. The relaxed atomic volume of BCC Np-U-Zr are tabulated in Table S2 and the atomic volume and volume of mixing are also plotted in Figure S2 of Supplementary Material. Following our parametrization of the effect of ion relaxation vs. atomic volume using the 10 BCC binary alloys formed between V, Nb, Ta, Mo and W[8], it is estimated based on the calculated atomic volume of elemental BCC Np, U and Zr that ion relaxation should only affect the mixing enthalpy of BCC Np-U-Zr negligibly, as having been discussed in detail for the Np-U system in Ref.[8]. Therefore, no corrections for neglecting ion relaxations is included when doing quantitative comparison of *ab initio* and CALHPAD enthalpies of mixing in this article.

## 3. Results and Discussion

### 3.1. Validation of CALPHAD model against experimental data

As explained in Introduction, the validation of our CALPHAD model in this section will focus on the BCC and liquid phases because for these high temperature phases reliable experimental data are available, at least in the binary subsystems, to allow meaningful assessment, and also because the use of no ternary interactions has been justified experimentally for these high temperature phases[9]. Results for the lower temperature phases are discussed only for the sake of completeness.

Let us start with the binary subsystems. The models for the U-Zr[5] and Np-Zr[7] binary subsystems have been validated in our previous work against existing experimental data, so here only the remaining binary subsystem Np-U is verified. Figure 3 shows the phase diagram of the Np-U system predicted by the CALPHAD model of this study and compares it to the experimental data of Mardon and Pearce[18]. The prediction from a previous CALPHAD model by Kurata[36] is also provided. As discussed in Ref.[8], Mardon and Pearce[18]'s study is the only one that probed the phase diagrarm of the Np-U system. Among their results, thermal analysis (black square) and dilatometry (orange circle) data points should be on the phase boundaries, while X-ray data only indicate whether the data points are in a single-phase (magenta triangle) or a two-phase (cyan start) field, and are in general not on any phase boundary. Figure 3 shows that for the high temperature liquid and BCC phases the phase boundaries are very well determined by thermal analysis data (dense black square data points)[18], and the two CALPHAD models also agree well and both reproduce these experimetnal data accurately. However, significant uncertainty exists for the low temperature part of the phase diagram, as discussed by us before[8]. Most available experimental data in that region are from X-ray that cannot unambiguously determine the phase boundaries. The two CALPHAD models also show significant differences there, especially near the Np rich end. It is further expected that low-temperature expermental data in general have a larger error bar than high-temperature data because of increased difficulty in obtaining equalibrated samples at lower temperature. As a result, no serious atempt is pursued to reproduce the experimental data, nor is it particularly meaniful to compare the two CALPHAD models for the low temepareture phase fields. Despite the limitations, our model for the Np-U binary system should be adquate to use together with our previous models for the U-Zr[5] and Np-Zr[7] binary systems to build a model for the Np-U-Zr ternary system, if the goal is for it to describe reasonably the high temperature BCC and liquid phases, which is discussed next.

Now let us compare our model for the full Np-U-Zr ternary system to experimental data. As mentioned in Introduction, to the best of our knowledge the study by Rodríguez *et al*.[9], despite performed over two decades ago, is still the only experimental phase diagram research on the Np-U-Zr ternary system to date. Figure 4 compares our model predicted isothermal sections at 793.15, 868.15 and 973.15 K to the experimental electron probe microanalysis and metallographic analysis data of Rodríguez et al.[9]. At 793.15 K shown in Figure 4 a), most of the experimental data are expected to reflect the equilibration of α(U) and δ phases as suggested by CALPHAD. However, the points that are expected to lie on the phase boundary of δ have $x$(U) that are about 0.1-0.3 larger than those predicted by CALPHAD. Such result that δ phase in Np-U-Zr is stable at $x$(U) smaller than 0.5 is surprising because it is not in line with the phase boundary of δ phase in the U-Zr system that has been determined rather accurately by many recent experimental studies, which our model for U-Zr reproduces quite well[5]. It seems unlikely that the addition of small content of Np would lead to such a substantially larger homogeneity range for δ. In fact, phase boundary of the δ phase of Np-Zr, which is also well measured in many recent experiments and reproduced in our model for Np-Zr[7], shows smaller homogeneity range than that for δ phase of U-Zr, further suggesting that δ Np-U-Zr should have smaller content of U than U-Zr as predicted in our model, not larger. On the other hand, those experimental points near the U end that likely correspond to α(U) have $x$(U) being 0.1 or more larger than those predicted by the CALPHAD model. However, Zr having very small solubility in α(U) is now rather well accepted—the solubility of Zr is found to be less than 0.02 in the entire stable temperature range of α(U), as reviewed in Ref.[5], so the discrepancy should again stem from the experimental side. The two remaining experimental points at 793.15 K in the middle of the triangle do not have any single-phase region nearby from CALPHAD but they may represent β(Zr) and γ(Np) as seen at 868.15 K in Figure 4 b). In fact, at this slightly higher temperature, although still showing some scattering, most of the experimental data suggest δ and α(U) phases to have substantially increased and decreased content of U, respectively, with boundaries of the two phases now being quite close to CALPHAD. Good agreement with our CALPHAD model may be represented by the two pairs of points in the middle of the triangle, which are very close to the β(Zr) and γ(Np) BCC phase boundaries at 868.15 K. Considering experimental data at 793.15 K and 868.15 together, large changes in $x$(U) (>0.1) for α(U) and δ with only 75 K difference in temperature seem unusual, especially the widening homogeneity range with increasing temperature. It is possible that data at one temperature are biased, likely those at 793.15 K, if assuming the CALPHAD model is reasonably accurate. Finally, going further up to 973.15 K, a good match with the remaining three pairs of experimental two-phase points cannot be found. However, Figure 4 c) shows that near the U-Zr side the two tie lines approximately cross at a single-phase point, which violates phase rule. Therefore, the data at 973.15 K are quite suspect. Altogether, the above analysis is in agreement with Rodríguez et al.'s comment[9] that due to the very small size of the phases identified, their "microanalysis values are relatively inaccurate". Based on the agreement with our CALPHAD model and tests of self-consistency, the data at 793.15 and 973.15 K in particular need further verification.

Rodríguez et al. [9] have also measured the melting temperatures of the same samples using dilatometer. However, melting point for multicomponent alloy is not unambiguously defined and depends on the direction of temperature change during the measurement. It actually refers to solidus temperature if heating and liquidus if cooling. Such distinction has not been made by Rodríguez et al. explicitly, although they have provided a dilatometric *heating* curve between 793.15 and 973.15 K and a comment that the *solidus* temperatures of Np-U-Zr are close to linear interpolation of those of the binary systems, which both suggest that what they have measured are probably solidus temperatures.

This supposition is further supported by the fact that our CALPHAD predicted solidus temperatures are much closer than liquidus to the melting temperatures that Rodríguez et al. reported in Ref.[9]. As Table 2 shows, except for sample R5, the differences in our model predicted solidus and the measured dilatometric melting temperatures are between 10 to 79 K with the average being 40 K, while the corresponding differences for liquidus temperatures are much larger, between 50 and 264 K with the average being 155 K. For sample R5, the solidus and liquidus temperatures from our model differ from the dilatometric melting temperature by 201 and 360 K, respectively. This sample may be affected by some unknown problem, because it is very close in compositions to other samples and it seems not very possible that the CALPHAD error suddenly surges at this single point. If assuming what Rodríguez et al.[9] have measured is indeed solidus temperatures and excluding sample R5, the agreement between our CALHPAD model and the experiment is satisfactory for an initial CALPHAD model developed based on limited experimental data.

Overall, to the extent possible with the limited data available, the above validations suggest that our model captures reasonably the phase stability of the high-temperature BCC and liquid phases of the Np-U-Zr system.

### 3.2. Transferability of $U_{\text{eff}}$ from binary to ternary

In this section CALPHAD predicted enthalpies are used as reference in lieu of experimental thermochemical data (mostly unavailable at present) to validate *ab initio* enthalpies for the BCC phase of the Np-U-Zr system. This means that our CALPHAD model is assumed to be reasonably accurate for the BCC phase—as supported by the comparisons to the experimental data of Rodríguez et al.[9] in last section—and is therefore taken as if it is exact to validate DFT and DFT + $U$ at different $U_{\text{eff}}$ parameters. Specifically, *ab initio* calculated enthalpy of mixing for the BCC phase is compared with that from CALPHAD. Such comparison is meaningful because errors in *ab initio* enthalpies have been found to be much larger than the uncertainty in CALPHAD for example in the U-Zr[5, 6] and Np-Zr[7] binary systems, for which our corresponding CALPHAD models should be quite robust thanks to the many reliable experimental data available.

The representative results are presented in Figure 5. A first observation is that the enthalpy surface from DFT (bottom left) is significantly higher than the CALPHAD one, most evidently on the Np-Zr rich side. The highest point of DFT enthalpy surface is near the point of $x$(Np)=0.75, $x$(U)=0, and $x$(Zr)=0.25 with a mixing enthalpy of 0.200 eV/atom (19.3 kJ/mole), while the corresponding CALPHAD value is 0.028 eV/atom (2.7 kJ/mole), about seven times smaller. The difference is smaller but still sizable on the U-Zr rich side, although relatively small on the Np-U side. Averaging over the 28 calculated compositions except for the three end points of BCC Np, U and Zr elemental metals, the root mean square (RMS) of the differences between DFT and CALPHAD enthalpy is 0.093 eV/atom (9.0 kJ/mole).

Now let us look at DFT + $U$ results. Navigating on the $U_{\text{eff}}$ grid first in the last row of Figure 5 where $U_{\text{eff}}$ (Np) is kept at 0 but $U_{\text{eff}}$ (U) is varied from 0 to 1.49 eV, the *ab initio* enthalpy surface not surprisingly adjusts lower on the U rich side but remains largely unchanged in the Np-rich end. The reverse is true in the first column where $U_{\text{eff}}$ (U) is kept at 0 but $U_{\text{eff}}$ (Np) is varied from 0 to 1.4 eV. If going along the diagonal the enthalpy drops on both the Np and the U sides. Overall, it seems that *ab initio* enthalpy becomes lower when both or either of $U_{\text{eff}}$(Np) and $U_{\text{eff}}$(Np) increase from 0 (i.e., only the standard DFT is applied) to nonzero values. However, when $U_{\text{eff}}$(Np) and $U_{\text{eff}}$(U) are too large (e.g., larger than 0.9 eV for the former and 1.24 for the later), the *ab initio* enthalpies either continues to reduce and becomes overly small or bounce back and become too large compared to CALPHAD. Moreover, near the Np-U side, the DFT enthalpy is already reasonably close to CALPHAD's. However, the change in enthalpy there is also quite small even when significant values of $U_{\text{eff}}$(Np) and $U_{\text{eff}}$(U) are used in DFT + $U$. Most encouragingly, when only one of $U_{\text{eff}}$(Np) and $U_{\text{eff}}$(U) is individually optimized to match CALPHAD for Np-Zr and/or U-Zr, enthalpy near the Np-U rich side also seems to be in general improved, although by smaller extent as the concentration of Zr decreases. The transferability of $U_{\text{eff}}$ values found here is also consistent with what is found in our recent study of the Np-U binary system. Therefore, a consistent set of $U_{eff}$ seems to apply to the BCC phase of the Np-U-Zr system in the whole ternary composition space.

A natural question to ask is then at what values of $U_{eff}$ (Np) and $U_{eff}$ (U) are enthalpies from CALPHAD and *ab initio* closest? This question is answered in Table 3, which presents RMS of the differences between the two enthalpies at the $U_{eff}$ grid we have illustrated in Figure 2. These actual *ab initio* calculated data in Table 3 are spline interpolated at $U_{eff}$ values not calculated and together they are visualized in Figure 6 as a three-dimensional (3D) surface of the RMS of differences plotted as a function of ($U_{eff}$ (Np), $U_{eff}$ (U)). It is clear from Figure 6 that a minimum exists at ($U_{eff}$ (Np), $U_{eff}$ (U)) ≈ (0.65, 0.99) eV, at which values the RMS of the differences in *ab initio* and CALPHAD enthalpies is 0.009 eV/atom (0.9 kJ/mole). ($U_{eff}$ (Np), $U_{eff}$ (U)) = (0.65, 0.99) should thus be the single-structure optimized $U_{eff}$ for the BCC phase in the Np-U-Zr ternary system. Note that the single-structure optimized $U_{eff}$'s for BCC phases in the U-Zr and Np-Zr binary systems determined individually in our previous studies of U-Zr and Np-Zr systems are also $U_{eff}$ (Np) = 0.65 eV and $U_{eff}$ (U) = 0.99 eV, so the binary and ternary systems have the same single-structure optimized $U_{eff}$(Np) and $U_{eff}$(Np). Moreover, when $U_{eff}$ for Np and U vary in the range of 0.65-0.9 and 1-1.5 eV, respectively, which are the ranges of single-structure optimized $U_{eff}$(Np) and $U_{eff}$(Np) determined individually in the U-Zr [6]and Np-Zr[7] binary systems for solid phases other than BCC, the average RMS differences remain 0.036 eV/atom (3.5 kJ/mole) or less, as given in Table 3. In particular, at ($U_{eff}$ (Np), $U_{eff}$ (U)) = (0.90, 1.24) eV, which are multi-structure optimized $U_{eff}$ (Np) and $U_{eff}$ (Np) determined in the Np-Zr and U-Zr binary systems, the error is about 0.026 eV/atom (2.5 kJ/mole).

Our finding that the same small ranges of $U_{eff}$ for Np and U can help improve the calculated enthalpy for Np and U based metallic actinide systems of different systems (e.g., the unary/binary/ternary; containing one or two actinides), compositions (e.g., across the whole BCC ternary in this study), and crystal structure (different solid phases) suggests that $U_{eff}$ should be predominately determined by the species and has reasonable transferability to different chemical and structural environments. This result implies that consistent or similar $U_{eff}$ determined in some benchmark metallic actinide systems may be applied to model alternative systems of different structure and composition, so long as the nature of the main chemical bonding characteristics remain similar to the validated systems. For example, the $U_{eff}$ values determined for Np and U are expected to be of value for modeling other metallic alloys of Np and U, although they should be quite different from those useful for actinides oxides due to significantly different chemical bonding characteristics, and future study to test this assertion is encouraged.

### 3.3. Predictions of the phase stability of Np-U-Zr at high temperature

After model validations in the previous two sections, this section presents and discusses phase stability for the high temperature BCC and liquid phases of the Np-U-Zr system predicted by the CALPHAD model, which is also cross-checked with results available from DFT + $U$ at ($U_{eff}$(Np), $U_{eff}$(U)) = (0.65, 0.99) eV—the single-structure optimized $U_{eff}$ for BCC Np-U-Zr discussed above.

#### 3.3.1. BCC Np-U-Zr solution behavior

Let us start with enthalpy of mixing for the BCC phase in Figure 7 from both CALPHAD and DFT + $U$. Viewing from all of the three Np-, U- and Zr-rich corners DFT+$U$ mixing enthalpy surface is very close to that of CALPHAD. Quantitatively, the magnitude of the mixing enthalpy of BCC Np-U-Zr is not very large in general—the maximum and average values of the *ab initio* calculated values are 0.058 and 0.029 eV/atom, respectively. The DFT+$U$ enthalpies are even slightly negative at $x$(Zr)=0 for Np-U, with minimum being -0.020 eV/atom (see the red ball symbol at the bottom of the cyan area in Figure 7 a) and b)), although the corresponding CALPHAD value is approximately zero. Negative values can still be consistent with the known phase stability and thermodynamics of this system, as discussed in Ref.[8]. In general, the following trends are noticed. The enthalpy is very small (<0.02 eV/atom) when Zr content is either high ($x$(Zr)>0.75) or very low ($x$(Zr)<0.1), which manifests in the enthalpy projection on the bottom plane as green area near the Zr end and the Np-U side; in the middle, however, the enthalpy increases to larger than 0.05 eV/atom and the projection there shows a large red area. Figure 7 c) also shows that in the middle region, the enthalpy is smaller when Np and U concentration is about the same. In fact, a saddle point seems to exist on the enthalpy surface near $x$(Np)= $x$(U)=0.375 and $x$(Zr)= 0.25.

Now let us look at the composition-temperature phase diagram of Np-U-Zr in three-dimensional in Figure 8, which shows only BCC and liquid phase boundaries (see also two-dimensional isothermal sections at 793.15, 868.15 and 973.15 K in Figure 4, and also at 900, 950, 1000 and 1050 K in Figure S3 of Supplementary Material). In Figure 8, the phase space is sectioned by three surfaces: liquidus, solidus and the BCC lower boundary. The first two are plotted in Figure 8 as color filled 3D surfaces, but the BCC lower boundary is outlined by curves made of color ball symbols—specifically, the red, black and the inner curled piece of green curves. Above the liquidus is the liquid phase; between liquidus and solidus is two-phase mixture of BCC and liquid; and above the BCC lower boundary and below the solidus is the BCC single phase. Below the BCC lower boundary are various phases that are equilibrating with BCC, in particular the BCC+ BCC' miscibility gap, the top boundary of which is outlined in Figure 8 by red and green curves. Figure 8 shows that 1) BCC+ BCC' miscibility gap does not show up when Zr content is either high or very low (i.e., near either the Zr end or the Np-U side); 2) a majority of the BCC+BCC' miscibility gap dissolves into a single BCC phase when temperature rises, however, a part of it near the Np-Zr side never becomes single BCC phase but melts into liquid directly from BCC+BCC', which is why an interface between solidus and BCC+BCC' miscibility gap curves exists. Such an interface is outlined by the two green curves.

The above phase diagram behavior can be understood in terms of the mixing enthalpy trends. In particular: 1) The Np-U rich side does not show stable BCC+BCC' miscibility gap, because the mixing enthalpy is very small, possibly even slightly negative. 2) The Np-Zr rich side does not show stable single BCC phase because Np melts before the BCC+BCC' miscibility gap converts to a single BCC phase due to entropy driven mixing. 3) The U-Zr rich side has similar positive mixing enthalpy as Np-Zr but the melting temperature of U is higher than Np, so the BCC+BCC' miscibility gap dissolves into a single BCC phase that is stable above about 1000 K until the solidus temperature, which will be discussed below.

### 3.3.2. Solidus and liquidus of Np-U-Zr

The solidus and liquidus temperatures of Np-U-Zr predicted by our CALHAD model are plotted in Figure 8. The liquidus temperature is very close to those expected from linear interpolations of the melting points of elemental Np, U and Zr metals, which is consistent with Rodríguez, *et al.*'s experimental data in Table 2. The solidus temperature surface is very close to the liquidus on the Np-U side, as shown in Figure 8 a), only slightly lower on the U-Zr side seen in Figure 8 b), but is significantly smaller on the Np-Zr side. As shown in Figure 8 c), in the whole region marked with green (i.e., the interface between solidus and BCC+BCC' miscibility gap), the solidus temperature stays essentially unchanged when $x$(Zr) increases from about 0.1 to 0.65. Even after $x$(Zr)>0.65, there is still a large difference between the liquidus and solidus. The two become much closer if the content of U is increased, for example when $x$(U)>0.5.

## 4. Conclusions

A CALPHAD model for the Np-U-Zr ternary system is developed based on Muggianu extrapolation of our models for the constituent binary systems. Comparing the model predicted isothermal sections and liquidus temperatures with available experimental data suggests that our model has reasonably characterized the high temperature part of the Np-U-Zr system. However, the lower temperature phases are uncertain due to lack of sufficient experimental data.

*Ab initio* calculations are performed at 28 compositions of the high-temperature BCC Np-U-Zr phase with both DFT and DFT + $U$. Referencing to our CALHAD model, DFT is found on average to overestimate the enthalpy of mixing for BCC Np-U-Zr by 0.093 eV/atom. DFT + $U$ predicts lower values than DFT and thus compares to CALPHAD more favorably using $U_{eff}$'s previously determined to give improved enthalpies in the Np-Zr and U-Zr binary systems that contain only one actinide element. Specifically, the error reduces to 0.026 eV/atom using multi-structure optimized $U_{eff}$'s (i.e., $U_{eff}$(Np), $U_{eff}$(U)) = (0.90, 1.24) eV) and between 0.009 and 0.036 eV/atom when $U_{eff}$(Np) and $U_{eff}$(U) vary in the ranges of single-structure optimized $U_{eff}$ (i.e., 0.65-0.9 and 1-1.5 eV for $U_{eff}$(Np) and $U_{eff}$(U), respectively). In particular, at ($U_{eff}$(Np,), $U_{eff}$(U))=(0.65, 0.99) eV, DFT + $U$ seems to consistently improve the enthalpy in the whole composition space of the BCC Np-U-Zr phase, resulting a minimal average difference from CALPHAD, 0.009 eV/atom (0.9 kJ/mole), suggesting that the single-structure optimized $U_{eff}$'s for BCC Np-U-Zr should also be $U_{eff}$(Np), $U_{eff}$(U)) = (0.65, 0.99) eV). Therefore, ternary BCC Np-U-Zr has the same single-structure optimized $U_{eff}$'s as binary BCC Np-Zr and BCC U-Zr.

Both CALPHAD and DFT + $U$ at BCC phase's single-structure optimized ($U_{eff}$(Np), $U_{eff}$(U)) = (0.65,0.99) eV predict enthalpy of mixing for BCC Np-U-Zr to be very small both near the Np-U side and on the Zr-rich end. In the intermediate region, mixing enthalpy is also not large when $x$(Np) and $x$(U) are about equal, and a saddle point near $x$(Np)=$x$(U)=0.375 and $x$(Zr)=0.25 exists. In U-Zr rich side, a BCC+BCC' miscibility gap evolves into a single BCC phase upon heating; near the Np-Zr side, however the miscibility gap does not turn into single BCC phase but instead melts directly when temperature increases due to the low melting point of Np. Near the Np-U side a miscibility gap does not exists due to the small mixing enthalpy. The liquidus of the Np-U-Zr system is close to linear interpolation of the melting points of the elemental metals, and the solidus is very close to the liquidus near the Np-U side, slightly lower near the U-Zr side, and significantly lower near the Np-Zr side. A large BCC+liquid two-phase region exists near the Zr end of the Np-Zr side.

Overall, this study has developed an initial quantitative thermodynamic model of the Np-U-Zr system that future work can continue to refine, found that DFT + $U$ can also improve the accuracy of enthalpy for the Np-U-Zr ternary systems with consistent $U_{eff}$'s from the binary subsystems, and provided some understanding on the solution behaviors of the BCC and liquid phases of the Np-U-Zr ternary alloy system. Based on the consistency in $U_{eff}$ between unary, binary and ternary systems of Np-U-Zr, it is believed that $U_{eff}$ determined in some benchmark metallic actinide systems can be transferable to alternative systems of different structure and composition, as long as they have similar chemical bonding characteristics. Further studies in more systems are needed to test this assertion.

## 5. Acknowledgement

Wei Xiong contributed to the CALPHAD modeling. Chao Jiang generated the $A_4B_3C_1$ ternary BCC SQS structure. Chuan Zhang assisted in multiple areas of this work. This research was funded by the U.S. Department of Energy Office of Nuclear Energy's Nuclear Energy University Programs under contract number 00088978. This work used the resources of the High Performance Computing Center at Idaho National Laboratory, which is supported by the Office of Nuclear Energy of the U.S. Department of Energy under Contract No. DE-AC07-05ID14517. This work also used the Extreme Science and Engineering Discovery Environment (XSEDE), which is supported by National Science Foundation grant number ACI-1053575. Thermo-Calc Software AB and CompuTherm, LLC are acknowledged for providing us the Thermo-Calc software and Pandat software, respectively.

# Tables

**Table 1**. Supercell structures and *k*-point meshes used in *ab initio* calculations of BCC Np-U-Zr.

| Supercell[1] | Number of compositions | Cell size (atoms/cell) | *k*-Point mesh |
|---|---|---|---|
| A | 3 | 1 | 17×17×17 |
| $A_3B_1$ | 6 | 16 | 6×6×6 |
| $A_1B_1$ | 3 | 16 | 6×6×6 |
| $A_2B_3C_3$ | 3 | 64 | 2×2×2 |
| $A_1B_1C_1$ | 1 | 36 | 3×3×3 |
| $A_2B_1C_1$ | 3 | 32 | 4×4×4 |
| $A_6B_1C_1$ | 3 | 64 | 2×2×2 |
| $A_4B_3C_1$ | 6 | 64 | 2×2×2 |

[1]A/B/C=Np, U, Zr

**Table 2**. Melting temperatures measured[9] at U-rich corner of Np-U-Zr compared with solidus and liquidus temperatures predicted from CALPHAD in this work.

| Sample[1] | x(Np) | x(U) | x(Zr) | Expt. (K) | CALPHAD (K) | |
|---|---|---|---|---|---|---|
| | | | | Melting $T$ | Solidus $T$ | Liquidus $T$ |
| R1  | 0.30 | 0.30 | 0.40 | 1243 | 1257 | 1507 |
| R2  | 0.11 | 0.67 | 0.22 | 1373 | 1436 | 1552 |
| R3  | 0.05 | 0.70 | 0.25 | 1443 | 1505 | 1614 |
| R4  | 0.05 | 0.80 | 0.15 | 1437 | 1447 | 1515 |
| R5  | 0.15 | 0.55 | 0.30 | 1228 | 1429 | 1588 |
| R6  | 0.20 | 0.70 | 0.10 | 1338 | 1328 | 1389 |
| R7  | 0.30 | 0.45 | 0.25 | 1267 | 1278 | 1438 |
| R9  | 0.16 | 0.60 | 0.24 | 1323 | 1400 | 1532 |
| R10 | 0.08 | 0.69 | 0.23 | 1388 | 1467 | 1572 |
| R11 | 0.10 | 0.80 | 0.10 | 1355 | 1390 | 1443 |

[1]Sample label and melting temperature from TABLE 3 of Rodriguez, *et al*.[9].

**Table 3**. Root mean square (RMS) of the differences in enthalpy of mixing between CALPHAD (300 K) and *ab initio* (0 K) for BCC Np-U-Zr calculated at the 28 compositions illustrated in Figure 1. DFT corresponds to the point at $U_{eff}$(Np)= $U_{eff}$(Np)=0 (bottom left), while DFT + $U$ to all others. Ranges of single-structure optimized $U_{eff}$'s are 0.65-0.90 for $U_{eff}$(Np)[7] and 0.99-1.49 eV for $U_{eff}$(U)[6], respectively. The units of enthalpy and $U_{eff}$ are eV/atom and eV, respectively.

| $U_{eff}$(U) \ $U_{eff}$(Np) | 0 | 0.15 | 0.4 | 0.65 | 0.9 | 1.15 | 1.4 |
|---|---|---|---|---|---|---|---|
| 1.49 | 0.073 | | 0.025 | | 0.036 | | 0.024 |
| 1.24 | | | | 0.016 | 0.026 | 0.018 | |
| 0.99 | 0.077 | | 0.020 | 0.009 | 0.017 | | 0.029 |
| 0.74 | | | | 0.014 | | | |
| 0.49 | 0.085 | | 0.032 | | 0.019 | | 0.047 |
| 0.24 | | 0.065 | | | | | |
| 0 | 0.093 | | 0.044 | | 0.032 | | 0.064 |

# Figures

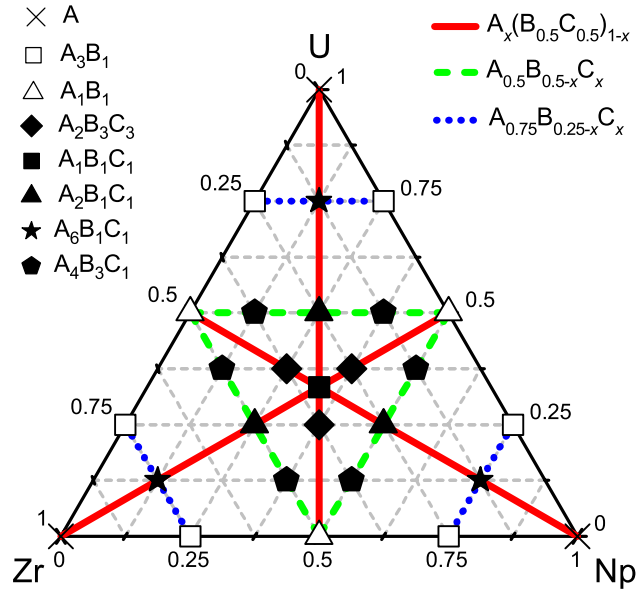

**Figure 1**. The 28 compositions (symbols) of BCC Np-U-Zr studied in *ab initio* calculations and the three representative series of isopleth paths (lines) that they form.

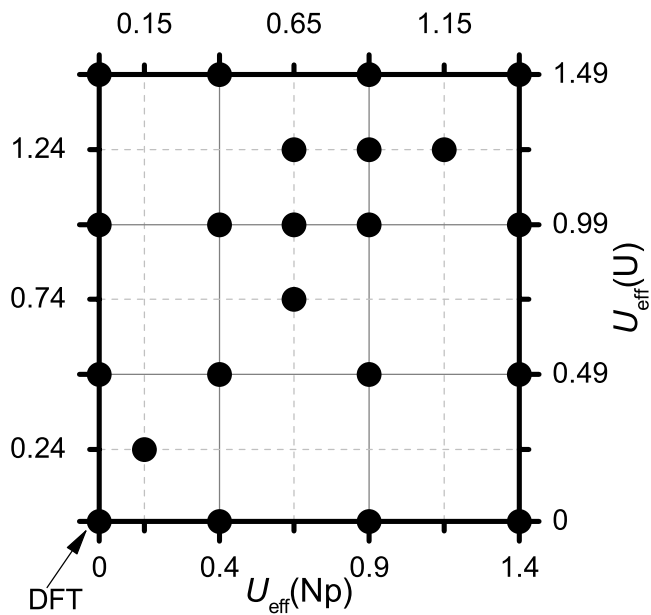

**Figure 2**. $U_{eff}$'s used for Np and U in DFT + $U$ calculations. DFT corresponds to the point at $U_{eff}(Np) = U_{eff}(Np) = 0$ (bottom left), while DFT + $U$ to all others.

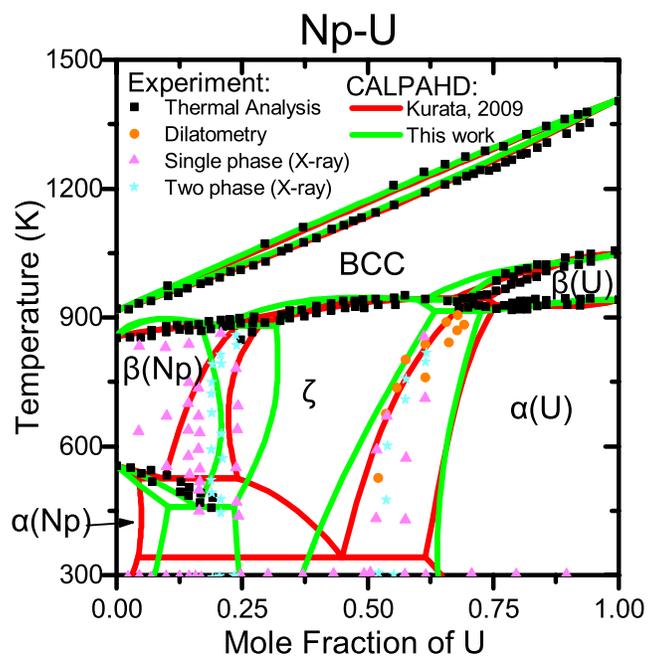

**Figure 3**. Phase diagram of the Np-U system from the CALPHAD models of Kurata[36] and this work compared to the experimental data of Mardon and Pearce[18].

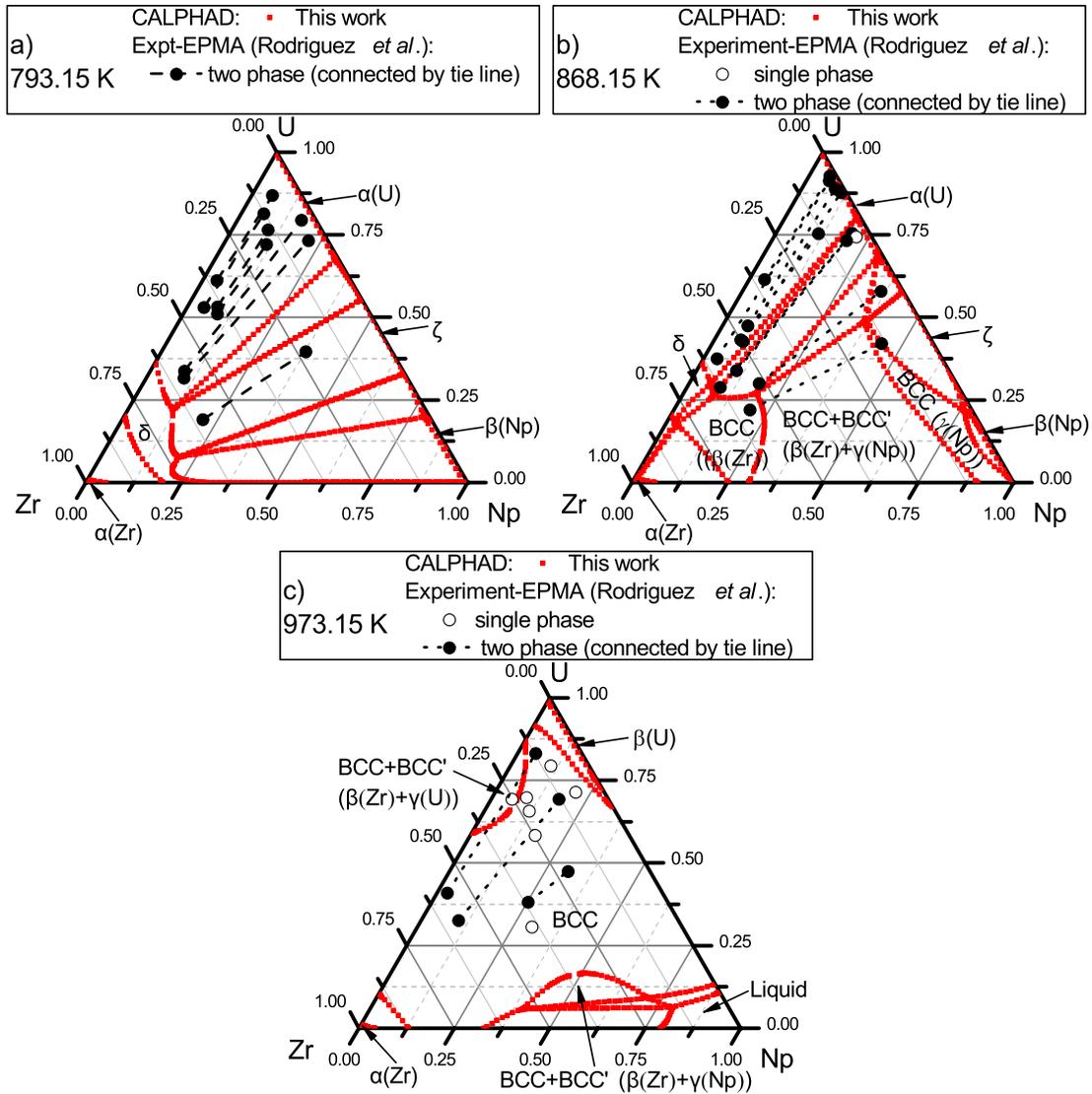

**Figure 4**. Isothermal sections of the phase diagram of Np-U-Zr at a) 793.15 K, b) 868.15 K, and c) 973.15 K. Greek phase labels for dominant component(s) of single-phase and BCC two-phase regions are given. Additionally, BCC and BCC+BCC' are used to denote BCC single- and two-phase regions, respectively.

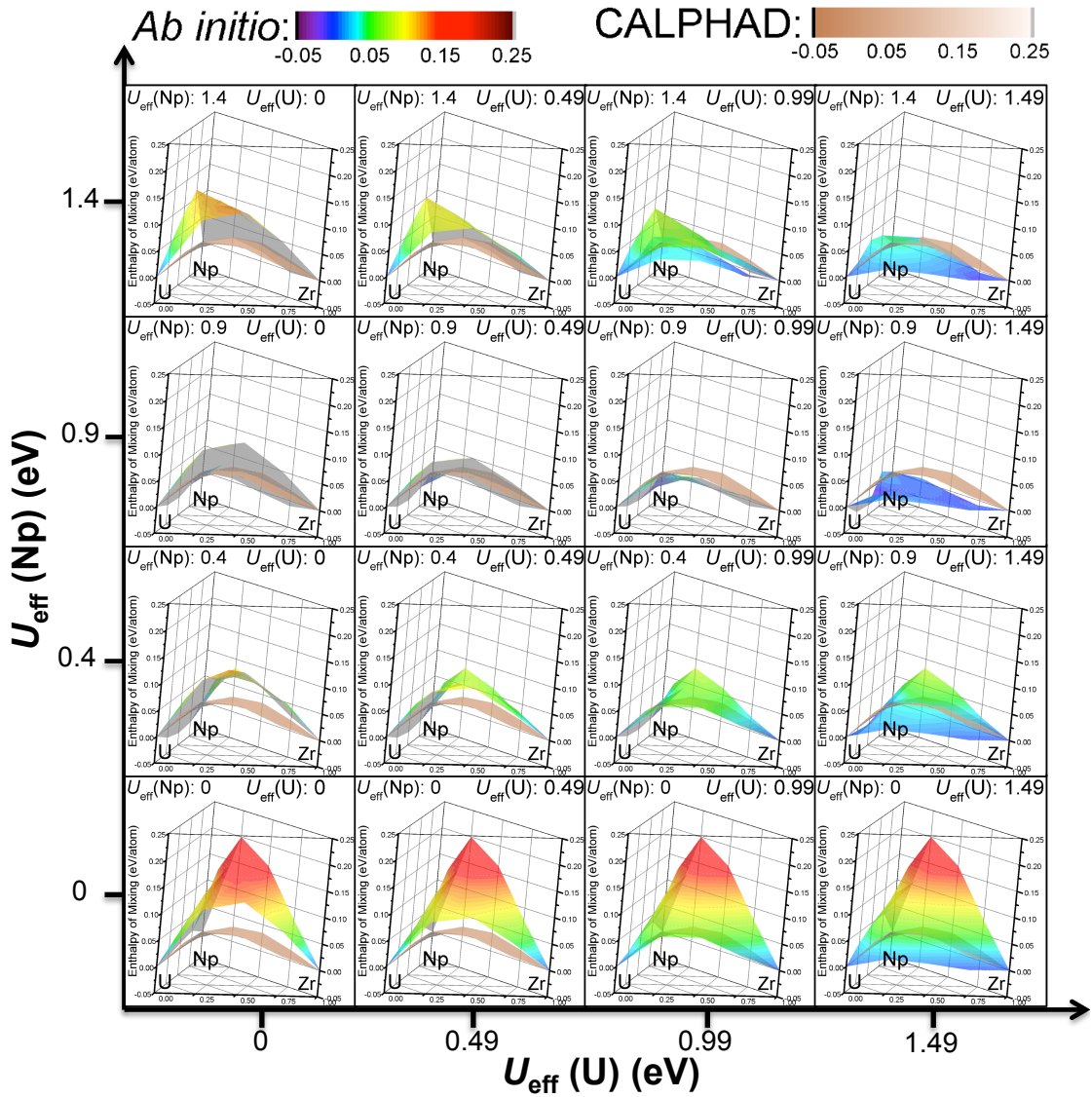

**Figure 5**. Enthalpy of mixing for BCC Np-U-Zr from CALPHAD (300 K) and *ab initio* (0 K) at different $U_{eff}$'s. DFT corresponds to the point at $U_{eff}(Np)= U_{eff}(Np)=0$ (bottom left), while DFT + $U$ to all others.

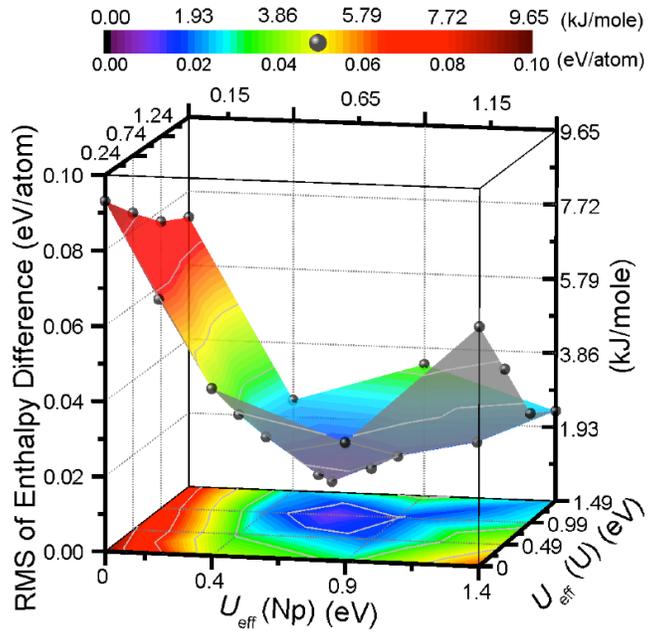

**Figure 6**. Root mean square (RMS) of the differences in enthalpy of mixing between CALPHAD (300 K) and *ab initio* (0 K) for the BCC phase of the Np-U-Zr system as a function of ($U_{eff}$(Np), $U_{eff}$(U)). DFT corresponds to the point at $U_{eff}$(Np)= $U_{eff}$(Np)=0 (bottom left), while DFT + $U$ to all others. *Ab initio* calculated values are marked black balls and the remaining in the surface are their spline interpolations. The bottom plane is a projection of the 3D surface.

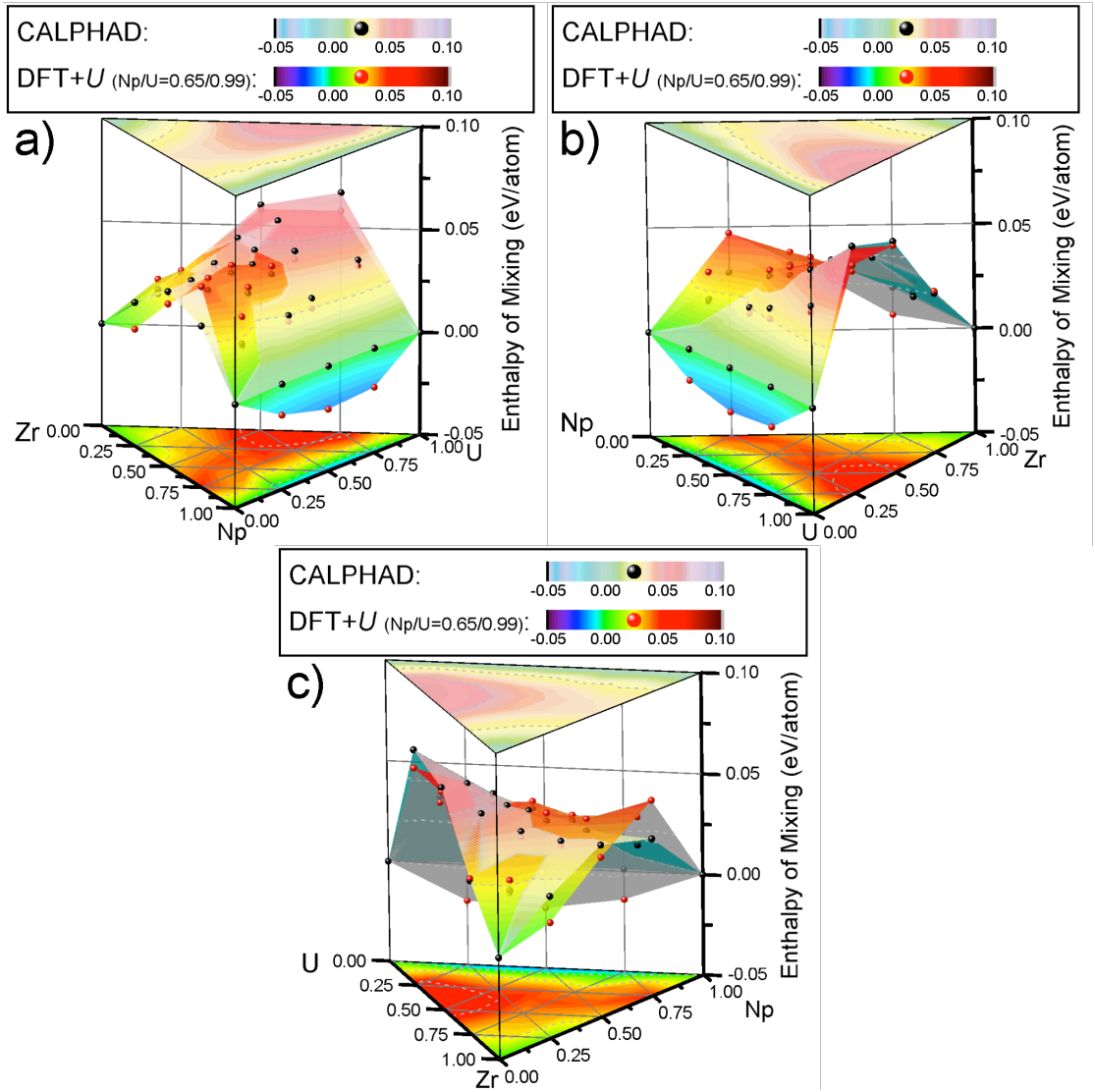

**Figure 7**. Enthalpy of mixing for the BCC phase of the Np-U-Zr system from CALPHAD and DFT + *U* calculated at ($U_{eff}$(Np), $U_{eff}$(U))=(0.6, 0.99) eV, viewed from a) Np-, b) U-, and c) Zr-rich corner. The front of CALPHAD and DFT + *U* surfaces are filled according to the color palettes given in the legend, while the back of them are filled by dark cyan and gray, respectively. The top and bottom planes are projections of the CALPHAD and DFT + *U* 3D surface, respectively. See **Figure 5** for the results from DFT.

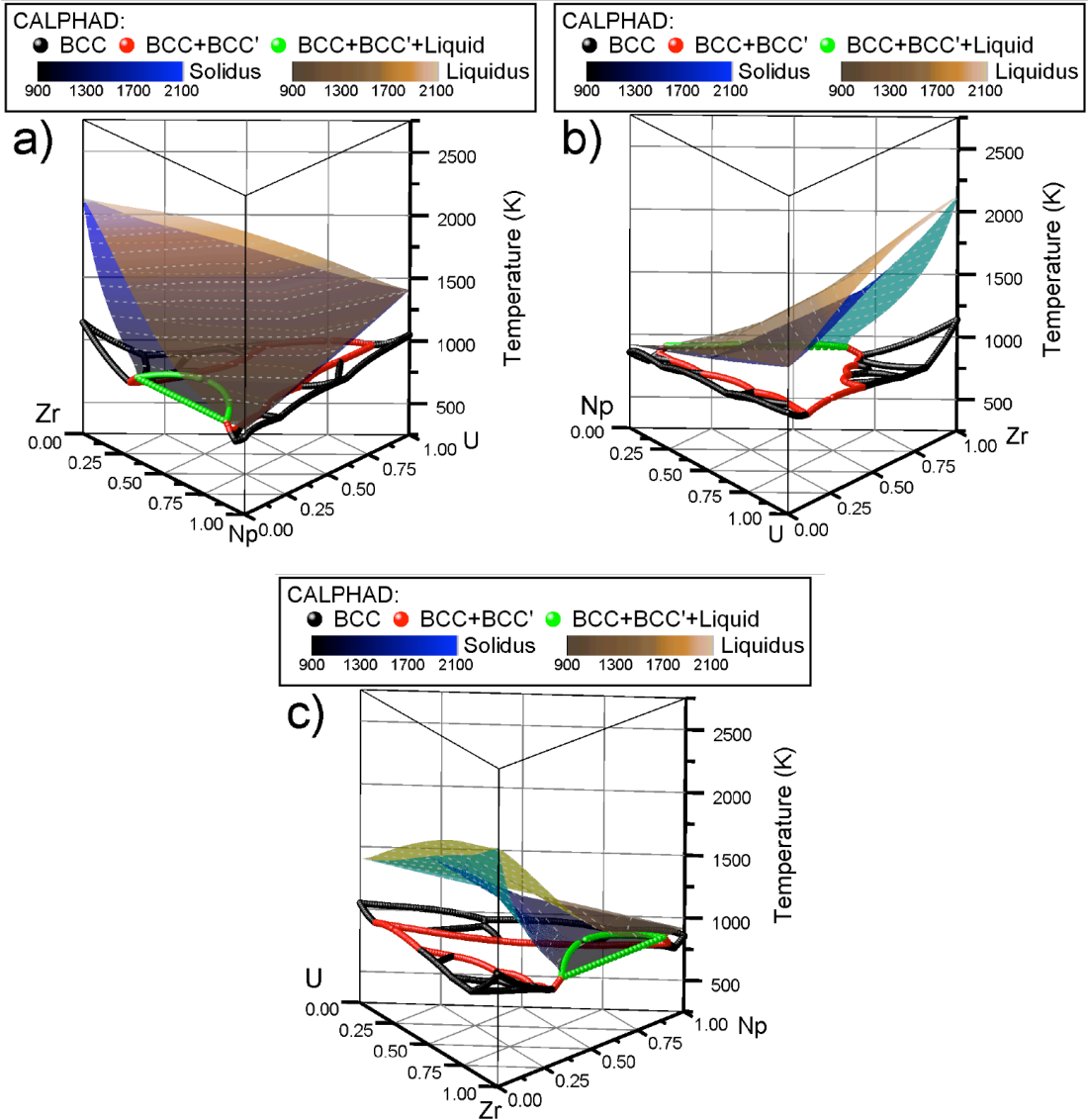

**Figure 8**. Phase diagram of for the Np-U-Zr system showing only the BCC and Liquid phases viewed from a) Np-, b) U-, and c) Zr-rich corner. Green curves outline the interface between solidus and BCC+BCC' miscibility gap. Black curves outline the interface between BCC and lower temperature phases. The front of solidus and liquidus surfaces are filled according to the color palettes given in the legend, while the back of them are filled by dark cyan and dark yellow, respectively. The surface outlined by red, black and green curves is the lower, while the surface of solidus is the upper boundary of the BCC single-phase region. The surface outlined by red and green curves is the upper boundary of BCC+BCC' miscibility gap.

# Supplemental material for
# "CALPHAD modeling and *ab initio* calculations of the Np-U-Zr system"

Wei Xie[1, a], Dane Morgan[1, b]

[1]Department of Materials Science and Engineering, University of Wisconsin-Madison, Madison WI 53706, United States

[a] Current address: Department of Materials Science and Engineering, University of California, Berkeley, Berkeley, CA 94720, United States
[b] Corresponding author. Address: 1509 University Avenue, Madison WI 53706, United States. Tel.: + 1 608 265 5879. Fax: +1 608 262 8353. Email address: ddmorgan@wisc.edu.

Table S1. CALPHAD model parameters in the Thermo-Calc Database (tdb) format for the Np-U-Zr ternary system from this work.

```
 ELEMENT /-   ELECTRON_GAS           0.0000E+00  0.0000E+00  0.0000E+00!
 ELEMENT VA   VACUUM                 0.0000E+00  0.0000E+00  0.0000E+00!
 ELEMENT NP   ORTHORHOMBIC_AC        2.3705E+02  0.0000E+00  0.0000E+00!
 ELEMENT U    ORTHORHOMBIC_A20       2.3803E+02  6.3640E+03  5.0200E+01!
 ELEMENT ZR   HCP_A3                 9.1224E+01  5.5663E+03  3.9181E+01!

 FUNCTION GLIQNP    298.15 -4627.18+160.024959*T-31.229*T*LN(T)
    -.0163885*T**2+2.941883E-06*T**3+439915*T**(-1); 917 Y
     -7415.255+247.671446*T-45.3964*T*LN(T); 4000 N !
 FUNCTION GBCCNP    298.15 -3224.664+174.911817*T-35.177*T*LN(T)
    -.00251865*T**2+5.14743E-07*T**3+302225*T**(-1); 856 Y
     -2366.486+180.807719*T-36.401*T*LN(T); 917 Y
     +50882.281-297.324358*T+30.7734*T*LN(T)-.0343483*T**2
     +2.707217E-06*T**3-7500100*T**(-1); 1999 Y
     -14879.686+254.773087*T-45.3964*T*LN(T); 4000 N !
 FUNCTION GHSERNP   298.15 +241.888-57.531347*T+4.0543*T*LN(T)
    -.04127725*T**2-402857*T**(-1); 553 Y
     -57015.112+664.27337*T-102.523*T*LN(T)+.0284592*T**2-2.483917E-06*T**3
     +4796910*T**(-1); 1799 Y
     -12092.736+255.780866*T-45.3964*T*LN(T); 4000 N !
 FUNCTION GHSERUU   298.15 -8407.734+130.955151*T-26.9182*T*LN(T)
    +.00125156*T**2-4.42605E-06*T**3+38568*T**(-1); 955 Y
     -22521.8+292.121093*T-48.66*T*LN(T); 3000 N !
 FUNCTION GLIQUU    298.15 +GHSERUU#+12355.5-10.3239*T; 3000 N !
 FUNCTION GHSERZR   130 -7827.595+125.64905*T-24.1618*T*LN(T)-.00437791*T**2
     +34971*T**(-1); 2128 Y
     -26085.921+262.724183*T-42.144*T*LN(T)-1.342896E+31*T**(-9); 6000 N !
 FUNCTION GLIQZR    298.15 +GHSERZR#+18147.69-9.080812*T+1.6275E-22*T**7;
    2128 Y
     -8281.26+253.812609*T-42.144*T*LN(T); 6000 N !
 FUNCTION GBCCUU    298.15 -752.767+131.5381*T-27.5152*T*LN(T)
    -.00835595*T**2+9.67907E-07*T**3+204611*T**(-1); 1049 Y
     -4698.365+202.685635*T-38.2836*T*LN(T); 3000 N !
 FUNCTION GBCCZR    298.15 -525.539+124.9457*T-25.607406*T*LN(T)
    -3.40084E-04*T**2-9.729E-09*T**3+25233*T**(-1)-7.6143E-11*T**4; 2128 Y
     -30705.955+264.284163*T-42.144*T*LN(T)+1.276058E+32*T**(-9); 6000 N !
 FUNCTION GHCPUU    298.15 +4247.233+131.5301*T-27.5152*T*LN(T)
    -.00835595*T**2+9.67907E-07*T**3+204611*T**(-1); 1049 Y
     +301.635+202.677635*T-38.2836*T*LN(T); 2500 N !
 FUNCTION GTETNP    298.15 -10157.32+183.829213*T-34.11*T*LN(T)
    -.0161186*T**2+4.98465E-06*T**3+532825*T**(-1); 555 Y
     -7873.688+207.01896*T-39.33*T*LN(T); 856 Y
     +19027.98-46.64846*T-3.4265*T*LN(T)-.01921045*T**2+1.52726E-06*T**3
     -3564640*T**(-1); 1999 Y
     -16070.82+256.707037*T-45.3964*T*LN(T); 4000 N !
 FUNCTION GTETUU    298.15 -5156.136+106.976316*T-22.841*T*LN(T)
    -.01084475*T**2+2.7889E-08*T**3+81944*T**(-1); 941.50 Y
```

```
       -14327.309+244.16802*T-42.9278*T*LN(T); 3000 N !
 FUNCTION GOMEGAZR  298.15 -8878.082+144.432234*T-26.8556*T*LN(T)
   -.002799446*T**2+38376*T**(-1); 2128 Y
    -29500.524+265.290858*T-42.144*T*LN(T)+7.17445E+31*T**(-9); 6000 N !
 FUNCTION UN_ASS    298.15 +0.0; 300 N !

 FUNCTION PARANPUB  298.15 6.3492187; 6000 N !
 FUNCTION PARANPUA  298.15 18073.114; 6000 N !

 TYPE_DEFINITION % SEQ *!
 DEFINE_SYSTEM_DEFAULT ELEMENT 2 !
 DEFAULT_COMMAND DEF_SYS_ELEMENT VA /- !

$ --------------------------------------------------------------------- Bcc_A2
 PHASE BCC_A2  %  2 1   3 !
   CONSTITUENT BCC_A2  :NP,U,ZR : VA :  !

  PARAMETER G(BCC_A2,NP:VA;0)         298.15 +GBCCNP#; 4000 N REF0 !
  PARAMETER G(BCC_A2,U:VA;0)          298.15 +GBCCUU#; 3000 N REF0 !
  PARAMETER G(BCC_A2,ZR:VA;0)         298.15 +GBCCZR#; 6000 N REF0 !

  PARAMETER G(BCC_A2,U,ZR:VA;0)       298.15 +23296.883-8.9731208*T;
   6000 N REF0 !
  PARAMETER G(BCC_A2,U,ZR:VA;1)       298.15 +21148.982-16.930263*T;
   6000 N REF0 !
  PARAMETER G(BCC_A2,U,ZR:VA;2)       298.15 2841.5860; 6000 N REF0 !

  PARAMETER G(BCC_A2,NP,ZR:VA;0)      298.15 +12333.36+3.9732*T;
   6000 N REFNPZR !
  PARAMETER G(BCC_A2,NP,ZR:VA;1)      298.15 +4304.17; 6000 N REFNPZR !

  PARAMETER G(BCC_A2,NP,U:VA;0)       298.15 5.7800000E+02; 3000 N REF0 !
$ --------------------------------------------------------------------- Delta
 PHASE DELTA  %  2 .666667   .333333 !
   CONSTITUENT DELTA  :NP,U,ZR : ZR :  !

  PARAMETER G(DELTA,U:ZR;0)           298.15  588.19+2.768*T
   +0.666667*GHSERUU+0.333333*GHSERZR; 6000 N REFUZR!
  PARAMETER G(DELTA,ZR:ZR;0)          298.15  527.50237+GHSERZR#;
   6000 N REF0 !
  PARAMETER G(DELTA,NP:ZR;0)          298.15  11173.871-14.3333*T
   +0.666667*GHSERNP#+0.333333*GHSERZR#;   6.00000E+03   N REFNPZR !

  PARAMETER L(DELTA,U,ZR:ZR;0)        298.15 -2209.76+6.740*T;
   6000 N REFUZR !
  PARAMETER L(DELTA,U,ZR:ZR;1)        298.15 236.69-5.874*T;
   6000 N REFUZR !
  PARAMETER G(DELTA,NP,ZR:ZR;0)       298.15 -26340.67+42.9988*T;
   6000 N REF0 !
  PARAMETER G(DELTA,NP,ZR:ZR;1)       298.15 -12443.65+15.8067*T;
   6000 N REF0 !
```

```
$ ------------------------------------------------------------------ Hcp_A3
 PHASE HCP_A3  %  2 1   .5 !
   CONSTITUENT HCP_A3  :NP,U,ZR : VA :  !

  PARAMETER G(HCP_A3,NP:VA;0)         298.15 +19000.0+GHSERNP#; 6000
   N REF0 !
  PARAMETER G(HCP_A3,U:VA;0)          298.15 +GHCPUU#; 2500 N REF0 !
  PARAMETER G(HCP_A3,ZR:VA;0)         298.15 +GHSERZR#; 6000 N REF0 !

  PARAMETER G(HCP_A3,U,ZR:VA;0)       298.15 24184.36; 6000 N REFUZR !
  PARAMETER G(HCP_A3,NP,ZR:VA;0)      298.15 -2109.32; 6000 N REF0 !

$ ------------------------------------------------------------------ LIQUID
 PHASE LIQUID  %  1  1.0  !
   CONSTITUENT LIQUID  :NP,U,ZR :  !

  PARAMETER G(LIQUID,NP;0)          298.15 +GLIQNP#; 4000 N REF0 !
  PARAMETER G(LIQUID,U;0)           298.15 +GLIQUU#; 3000 N REF0 !
  PARAMETER G(LIQUID,ZR;0)          298.15 +GLIQZR#; 6000 N REF0 !

  PARAMETER G(LIQUID,U,ZR;0)        298.15  33465.24-14.555*T;
   6.00000E+03  N REFUZR !
  PARAMETER G(LIQUID,U,ZR;1)        298.15  19809.38-18.068*T;
   6.00000E+03  N REFUZR !
  PARAMETER G(LIQUID,NP,ZR;0)       298.15  1142.97; 6000 N REFNPZR !
  PARAMETER G(LIQUID,NP,ZR;1)       298.15  10193.88; 6000 N REFNPZR !

$ ------------------------------------------------------------------ OMEGA
$ PHASE OMEGA  %  1  1.0  !
$   CONSTITUENT OMEGA  :ZR :  !

$   PARAMETER G(OMEGA,ZR;0)           298.15 +GOMEGAZR#; 6000 N REF0 !

$ ------------------------------------------------------------ ORTHORHOMBIC_A20
$                              ALPHA_U
 PHASE ORTHORHOMBIC_A20  %  1  1.0  !
   CONSTITUENT ORTHORHOMBIC_A20  :NP,U,ZR :  !

  PARAMETER G(ORTHORHOMBIC_A20,NP;0)    298.15 11178.000+GHSERNP#; 4000
   N REF0 !
  PARAMETER G(ORTHORHOMBIC_A20,U;0)     298.15 +GHSERUU#; 3000 N REF0 !
  PARAMETER G(ORTHORHOMBIC_A20,ZR;0)    298.15 +4474.461+124.9457*T
 -25.607406*T*LN(T)-3.40084E-04*T**2-9.729E-09*T**3+25233*T**(-1)
 -7.6143E-11*T**4; 2128 Y
  -25705.955+264.284163*T-42.144*T*LN(T)+1.276058E+32*T**(-9); 6000 N REF0 !
  PARAMETER G(ORTHORHOMBIC_A20,U,ZR;0)  298.15 3.0312439E+04; 6000 N REF0 !
  PARAMETER G(ORTHORHOMBIC_A20,NP,U;0)  298.15 -27917.331+7.6772017*T;
   3000 N REF0 !
  PARAMETER G(ORTHORHOMBIC_A20,NP,U;1)  298.15 -12441.161+3.8426518*T;
   3000 N REF0 !
```

```
$ ------------------------------------------------------------- TETRAGONAL_U
$                                    BETA_U
 PHASE TETRAGONAL_U  %  1  1.0  !
   CONSTITUENT TETRAGONAL_U  :NP,U,ZR :  !

   PARAMETER G(TETRAGONAL_U,NP;0)      298.15 13579.0+GTETNP#;
    6000    N REF0 !
   PARAMETER G(TETRAGONAL_U,U;0)       298.15 -5156.136+106.976316*T
  -22.841*T*LN(T)-.01084475*T**2+2.7889E-08*T**3+81944*T**(-1); 941.50 Y
   -14327.309+244.16802*T-42.9278*T*LN(T); 3000 N REF0 !
   PARAMETER G(TETRAGONAL_U,ZR;0)      298.15 +4474.461+124.9457*T
  -25.607406*T*LN(T)-3.40084E-04*T**2-9.729E-09*T**3+25233*T**(-1)
  -7.6143E-11*T**4; 2128 Y
   -25705.955+264.284163*T-42.144*T*LN(T)+1.276058E+32*T**(-9); 6000 N REF0 !
   PARAMETER G(TETRAGONAL_U,U,ZR;0)    298.15 2.7980588E+04; 6000 N REF0 !
   PARAMETER G(TETRAGONAL_U,NP,U;0)    298.15 -20142.460-2.1277030*T;
    3000 N REF0 !
   PARAMETER G(TETRAGONAL_U,NP,U;1)    298.15 -10101.902; 3000 N REF0 !

$ ------------------------------------------------------------- ORTHORHOMBIC_AC
$                                    ALPHA_NP
 PHASE ORTHORHOMBIC_AC  %  1  1.0  !
   CONSTITUENT ORTHORHOMBIC_AC  :NP,U,ZR :  !

   PARAMETER G(ORTHORHOMBIC_AC,NP;0)    298.15 +GHSERNP#; 4000 N REF0 !
   PARAMETER G(ORTHORHOMBIC_AC,U;0)     298.15 4266.0000+4.1605715E-01*T
  +GHSERUU#; 4000   N REF0 !
   PARAMETER G(ORTHORHOMBIC_AC,ZR;0)    298.15 +4474.461+124.9457*T
  -25.607406*T*LN(T)-3.40084E-04*T**2-9.729E-09*T**3+25233*T**(-1)
  -7.6143E-11*T**4; 2128 Y
   -25705.955+264.284163*T-42.144*T*LN(T)+1.276058E+32*T**(-9); 6000 N REF0 !
   PARAMETER G(ORTHORHOMBIC_AC,NP,ZR;0)  298.15 528764.83; 6000 N REF0 !
   PARAMETER G(ORTHORHOMBIC_AC,NP,U;0)   298.15 -10413.603+1.0022274*T;
    4000 N REF0 !

$ ------------------------------------------------------------- TETRAGONAL_AD
$                                    BETA_NP
 PHASE TETRAGONAL_AD  %  1  1.0  !
   CONSTITUENT TETRAGONAL_AD  :NP,U,ZR :  !

   PARAMETER G(TETRAGONAL_AD,NP;0)     298.15 -10157.32+183.829213*T
  -34.11*T*LN(T)-.0161186*T**2+4.98465E-06*T**3+532825*T**(-1); 555 Y
   -7873.688+207.01896*T-39.33*T*LN(T); 856 Y
   +19027.98-46.64846*T-3.4265*T*LN(T)-.01921045*T**2+1.52726E-06*T**3
  -3564640*T**(-1); 1999 Y
   -16070.82+256.707037*T-45.3964*T*LN(T); 4000 N REF0 !
   PARAMETER G(TETRAGONAL_AD,U;0)      298.15 11420.00+2.2060391*T+GTETUU#;
    6000    N REF0 !
   PARAMETER G(TETRAGONAL_AD,ZR;0)     298.15 +4474.461+124.9457*T
  -25.607406*T*LN(T)-3.40084E-04*T**2-9.729E-09*T**3+25233*T**(-1)
```

```
    -7.6143E-11*T**4; 2128 Y
    -25705.955+264.284163*T-42.144*T*LN(T)+1.276058E+32*T**(-9); 6000 N REF0 !
   PARAMETER G(TETRAGONAL_AD,NP,ZR;0)    298.15 23559.89; 6000 N REFNPZR !
   PARAMETER G(TETRAGONAL_AD,NP,U;0)    298.15 -40533.601+28.627404*T;
   3000 N REF0 !
$ ------------------------------------------------------------------ THETA
 PHASE THETA % 2 4  1 !
   CONSTITUENT THETA  :NP : ZR : !

   PARAMETER G(THETA,NP:ZR;0)        298.15 -2076.74+4*GHSERNP#
   +GHSERZR#; 6000 N REF0 !

$ ------------------------------------------------------------------ ZETA

 PHASE ZETA % 2 1  2 !
   CONSTITUENT ZETA  :NP,U : NP,U : !

   PARAMETER G(ZETA,NP:NP;0)         298.15 9508.2711+3*GHSERNP#;
   3000 N REF0 !
   PARAMETER G(ZETA,U:NP;0)          298.15 -7622.1785-6.3492187*T
   +GHSERUU#+2*GHSERNP#; 3000 N REF0 !
   PARAMETER G(ZETA,NP:U;0)          298.15 +PARANPUA#+PARANPUB#*T
   +GHSERNP#+2*GHSERUU#; 3000 N REF0 !
   PARAMETER G(ZETA,U:U;0)           298.15 9.4626644E+02+3*GHSERUU#;
   3000 N REF0 !
   PARAMETER G(ZETA,NP,U:NP;0)       298.15 -11701.428-4.1083552*T;
   6000 N REF0 !
   PARAMETER G(ZETA,NP:NP,U;0)       298.15 -11701.428-4.1083552*T;
   6000 N REF0 !
   PARAMETER G(ZETA,U:NP,U;0)        298.15 6.8717203E+03-1.2098479E-01*T;
   6000 N REF0 !
   PARAMETER G(ZETA,NP,U:U;0)        298.15 6.8717203E+03-1.2098479E-01*T;
   6000 N REF0 !

 LIST_OF_REFERENCES
 NUMBER  SOURCE
   REF0   'Unindex reference'
   REF1   'PURE1 - SGTE Pure Elements (Unary) Database (Version 1.0),
       developed by SGTE (Scientific Group Thermodata Europe), 1991-1992,
       and provided by TCSAB (Jan. 1991). Also in: Dinsdale A. (1991):
       SGTE data for pure elements, Calphad, 15, 317-425.'
   REF3   'PURE3 - SGTE Pure Elements (Unary) Database (Version 3.0),
       developed by SGTE (Scientific Group Thermodata Europe), 1991-1996,
       and provided by TCSAB (Aug. 1996). '
   REFUZR  'U-ZR by Wei Xiong, UW-Madison, J Nulcear Mater., 2013.'
   REFNPZR 'NP-ZR by Wei Xiong, UW-Madison, 2013.'
   REF4   'PURE4 - SGTE Pure Elements (Unary) Database (Version 4.6),
       developed by SGTE (Scientific Group Thermodata Europe), 1991-2008,
       and provided by TCSAB (Jan. 2008).
```

Table S2. Ternary SQS supercells in POSCAR format of VASP used in this study.

| Label | Structure |
|---|---|
| A1B1C1 | A1B1C1<br>3.0<br>2.000000 -2.000000 0.000000<br>-2.000000 0.000000 2.000000<br>-1.500000 -1.500000 -1.500000<br>12 12 12<br>Direct<br>0.833333 0.166666 0.222222 A<br>0.833333 0.166666 0.555556 A<br>0.333334 0.166666 0.222222 A<br>0.833333 0.666667 0.222222 A<br>0.500000 0.500000 0.666667 A<br>1.000000 1.000000 0.666667 A<br>0.166666 0.333334 0.777778 A<br>1.000000 1.000000 1.000000 A<br>0.666667 0.833333 0.111111 A<br>0.833333 0.666667 0.888889 A<br>1.000000 1.000000 0.333333 A<br>0.500000 1.000000 1.000000 A<br>0.166666 0.333334 0.111111 B<br>0.333334 0.666667 0.222222 B<br>0.666667 0.333334 0.444444 B<br>1.000000 0.500000 0.666667 B<br>0.500000 0.500000 0.333333 B<br>0.666667 0.333334 0.111111 B<br>0.333334 0.166666 0.555556 B<br>0.166666 0.333334 0.444444 B<br>0.333334 0.166666 0.888889 B<br>0.166666 0.833333 0.777778 B<br>0.166666 0.833333 0.444444 B<br>0.166666 0.833333 0.111111 B<br>0.666667 0.833333 0.444444 C<br>1.000000 0.500000 1.000000 C<br>0.333334 0.666667 0.888889 C<br>0.666667 0.333334 0.777778 C<br>0.500000 1.000000 0.333333 C<br>1.000000 0.500000 0.333333 C<br>0.500000 0.500000 1.000000 C<br>0.500000 1.000000 0.666667 C<br>0.833333 0.166666 0.888889 C<br>0.333334 0.666667 0.555556 C<br>0.666667 0.833333 0.777778 C<br>0.833333 0.666667 0.555556 C |

| A2B1C1 | A2B1C1 |
|---|---|
| | 3.0 |
| | 1.000000 -1.000000 3.000000 |
| | 1.000000 3.000000 -1.000000 |
| | -3.000000 -1.000000 -1.000000 |
| | 16 8 8 |
| | Direct |
| | 0.250000 0.500000 0.750000 A |
| | 0.750000 1.000000 0.250000 A |
| | 0.750000 1.000000 0.750000 A |
| | 0.500000 0.250000 0.250000 A |
| | 0.500000 0.500000 0.500000 A |
| | 0.500000 0.500000 1.000000 A |
| | 0.500000 1.000000 1.000000 A |
| | 0.750000 0.750000 0.500000 A |
| | 0.750000 0.750000 1.000000 A |
| | 0.250000 1.000000 0.750000 A |
| | 0.500000 0.750000 0.750000 A |
| | 1.000000 0.250000 0.250000 A |
| | 1.000000 0.750000 0.750000 A |
| | 0.500000 1.000000 0.500000 A |
| | 0.250000 0.250000 1.000000 A |
| | 0.750000 0.250000 1.000000 A |
| | 0.750000 0.500000 0.750000 B |
| | 0.500000 0.750000 0.250000 B |
| | 0.250000 0.750000 0.500000 B |
| | 0.250000 0.750000 1.000000 B |
| | 0.750000 0.500000 0.250000 B |
| | 1.000000 0.250000 0.750000 B |
| | 1.000000 0.500000 0.500000 B |
| | 1.000000 0.500000 1.000000 B |
| | 0.250000 1.000000 0.250000 C |
| | 1.000000 1.000000 1.000000 C |
| | 0.750000 0.250000 0.500000 C |
| | 0.500000 0.250000 0.750000 C |
| | 1.000000 1.000000 0.500000 C |
| | 0.250000 0.250000 0.500000 C |
| | 0.250000 0.500000 0.250000 C |
| | 1.000000 0.750000 0.250000 C |

| A2B3C3 | A2B3C3 |
|---|---|
| | A2B3C3<br>3.0<br>2.000000 2.000000 -2.000000<br>-2.000000 -2.000000 -2.000000<br>-2.000000 2.000000 2.000000<br>16 24 24<br>Direct<br>0.500000 0.750000 0.750000 A<br>1.000000 0.500000 0.750000 A<br>0.750000 0.250000 0.250000 A<br>0.750000 0.750000 0.250000 A<br>0.500000 0.500000 0.750000 A<br>0.750000 0.250000 0.750000 A<br>0.750000 0.500000 0.750000 A<br>0.500000 0.500000 1.000000 A<br>0.250000 0.250000 0.500000 A<br>0.750000 0.750000 0.750000 A<br>0.250000 1.000000 0.250000 A<br>0.750000 1.000000 0.250000 A<br>1.000000 0.250000 0.500000 A<br>0.500000 1.000000 1.000000 A<br>0.250000 0.750000 1.000000 A<br>0.750000 1.000000 0.500000 A<br>1.000000 0.750000 0.250000 B<br>1.000000 0.500000 0.250000 B<br>0.750000 1.000000 0.750000 B<br>1.000000 0.500000 0.500000 B<br>1.000000 0.500000 1.000000 B<br>1.000000 1.000000 1.000000 B<br>0.750000 0.250000 1.000000 B<br>0.250000 0.500000 0.500000 B<br>0.750000 0.500000 0.500000 B<br>1.000000 0.250000 0.750000 B<br>0.500000 0.500000 0.250000 B<br>0.250000 0.750000 0.250000 B<br>0.250000 0.750000 0.750000 B<br>0.250000 0.500000 0.750000 B<br>1.000000 0.750000 0.750000 B<br>1.000000 1.000000 0.250000 B<br>1.000000 1.000000 0.750000 B<br>0.250000 0.250000 0.750000 B<br>0.250000 0.500000 0.250000 B<br>1.000000 0.750000 0.500000 B<br>0.750000 0.250000 0.500000 B<br>0.750000 0.750000 1.000000 B<br>0.250000 1.000000 1.000000 B<br>0.750000 0.500000 1.000000 B<br>0.500000 0.250000 0.500000 C<br>1.000000 0.750000 1.000000 C<br>0.250000 0.500000 1.000000 C<br>0.500000 0.750000 0.250000 C<br>1.000000 0.250000 0.250000 C<br>0.500000 1.000000 0.250000 C<br>0.500000 1.000000 0.750000 C<br>0.250000 1.000000 0.750000 C<br>0.750000 0.500000 0.250000 C<br>0.500000 0.750000 0.500000 C<br>0.500000 0.750000 1.000000 C<br>0.500000 0.500000 0.500000 C<br>0.500000 1.000000 0.500000 C<br>0.250000 0.250000 1.000000 C<br>0.250000 0.750000 0.500000 C<br>0.250000 1.000000 0.500000 C<br>0.750000 1.000000 1.000000 C<br>0.500000 0.250000 0.250000 C<br>0.500000 0.250000 0.750000 C<br>0.250000 0.250000 0.250000 C<br>0.500000 0.250000 1.000000 C<br>1.000000 0.250000 1.000000 C<br>1.000000 1.000000 0.500000 C<br>0.750000 0.750000 0.500000 C |

| A6B1C1 | A6B1C1 |
|---|---|
| | 3.0 |
| | 2.000000 2.000000 -2.000000 |
| | -2.000000 -2.000000 -2.000000 |
| | -2.000000 2.000000 2.000000 |
| | 48 8 8 |
| | Direct |
| | 0.500000 0.250000 0.250000 A |
| | 1.000000 0.250000 0.250000 A |
| | 0.500000 1.000000 0.250000 A |
| | 1.000000 0.500000 0.750000 A |
| | 1.000000 1.000000 0.250000 A |
| | 1.000000 1.000000 0.750000 A |
| | 0.250000 0.750000 0.750000 A |
| | 0.750000 0.750000 0.250000 A |
| | 0.750000 0.750000 0.750000 A |
| | 0.250000 0.500000 0.250000 A |
| | 0.750000 0.500000 0.750000 A |
| | 0.500000 0.250000 1.000000 A |
| | 0.500000 0.750000 1.000000 A |
| | 1.000000 0.250000 0.500000 A |
| | 1.000000 0.750000 1.000000 A |
| | 0.500000 1.000000 0.500000 A |
| | 1.000000 0.500000 1.000000 A |
| | 1.000000 1.000000 1.000000 A |
| | 0.250000 0.750000 0.500000 A |
| | 0.750000 0.250000 1.000000 A |
| | 0.750000 0.500000 0.500000 A |
| | 0.750000 1.000000 0.500000 A |
| | 1.000000 0.750000 0.250000 A |
| | 0.500000 0.500000 0.750000 A |
| | 1.000000 0.500000 0.250000 A |
| | 0.250000 0.250000 0.250000 A |
| | 0.250000 1.000000 0.250000 A |
| | 0.500000 1.000000 1.000000 A |
| | 1.000000 0.750000 0.750000 A |
| | 0.750000 0.750000 1.000000 A |
| | 0.750000 0.500000 0.250000 A |
| | 0.750000 1.000000 0.750000 A |
| | 0.500000 0.750000 0.750000 A |
| | 0.500000 0.500000 0.250000 A |
| | 0.500000 0.750000 0.500000 A |
| | 0.500000 0.750000 0.250000 A |
| | 1.000000 0.250000 0.750000 A |
| | 0.250000 0.250000 1.000000 A |
| | 0.750000 0.250000 0.500000 A |
| | 0.250000 1.000000 1.000000 A |
| | 0.250000 0.750000 1.000000 A |
| | 1.000000 0.250000 1.000000 A |
| | 0.250000 0.750000 0.250000 A |
| | 0.750000 1.000000 0.250000 A |
| | 0.250000 0.500000 1.000000 A |
| | 0.250000 1.000000 0.500000 A |
| | 0.250000 0.250000 0.500000 A |
| | 1.000000 0.750000 0.500000 A |
| | 0.250000 0.250000 0.750000 B |
| | 0.750000 0.250000 0.750000 B |
| | 0.750000 0.750000 0.500000 B |
| | 0.250000 0.500000 0.500000 B |
| | 1.000000 1.000000 0.500000 B |
| | 0.500000 1.000000 0.750000 B |
| | 1.000000 0.500000 0.500000 B |
| | 0.250000 1.000000 0.750000 B |
| | 0.500000 0.250000 0.500000 C |
| | 0.500000 0.500000 0.500000 C |
| | 0.500000 0.500000 1.000000 C |
| | 0.500000 0.250000 0.750000 C |
| | 0.750000 0.250000 0.250000 C |
| | 0.750000 0.500000 1.000000 C |
| | 0.750000 1.000000 1.000000 C |
| | 0.250000 0.500000 0.750000 C |

| A4B3C1 | A4B3C1 |
|---|---|
| | 3.60 |
| | 2.000000 2.000000 2.000000 |
| | 2.000000 -2.000000 2.000000 |
| | 2.000000 2.000000 -2.000000 |
| | 32 24 8 |
| | Direct |
| | 0.250000 0.000000 0.750000 A |
| | 0.250000 0.250000 0.500000 A |
| | 0.250000 0.500000 0.500000 A |
| | 0.250000 0.500000 0.750000 A |
| | 0.250000 0.750000 0.250000 A |
| | 0.500000 0.250000 0.000000 A |
| | 0.500000 0.250000 0.250000 A |
| | 0.500000 0.500000 0.750000 A |
| | 0.500000 0.750000 0.250000 A |
| | 0.500000 0.750000 0.500000 A |
| | 0.750000 0.000000 0.250000 A |
| | 0.750000 0.000000 0.500000 A |
| | 0.750000 0.000000 0.750000 A |
| | 0.750000 0.250000 0.250000 A |
| | 0.750000 0.750000 0.750000 A |
| | 0.000000 0.000000 0.000000 A |
| | 0.000000 0.000000 0.500000 A |
| | 0.000000 0.250000 0.250000 A |
| | 0.000000 0.750000 0.250000 A |
| | 0.250000 0.000000 0.000000 A |
| | 0.250000 0.000000 0.250000 A |
| | 0.500000 0.000000 0.500000 A |
| | 0.500000 0.500000 0.500000 A |
| | 0.500000 0.750000 0.000000 A |
| | 0.750000 0.000000 0.000000 A |
| | 0.750000 0.250000 0.500000 A |
| | 0.750000 0.250000 0.750000 A |
| | 0.750000 0.750000 0.250000 A |
| | 0.000000 0.500000 0.750000 A |
| | 0.250000 0.500000 0.000000 A |
| | 0.500000 0.000000 0.250000 A |
| | 0.500000 0.250000 0.500000 A |
| | 0.000000 0.000000 0.750000 B |
| | 0.000000 0.750000 0.500000 B |
| | 0.250000 0.250000 0.000000 B |
| | 0.250000 0.500000 0.250000 B |
| | 0.250000 0.750000 0.000000 B |
| | 0.250000 0.750000 0.750000 B |
| | 0.500000 0.250000 0.750000 B |
| | 0.500000 0.500000 0.000000 B |
| | 0.500000 0.500000 0.250000 B |
| | 0.750000 0.500000 0.500000 B |
| | 0.750000 0.750000 0.000000 B |
| | 0.000000 0.250000 0.750000 B |
| | 0.000000 0.500000 0.250000 B |
| | 0.000000 0.750000 0.000000 B |
| | 0.000000 0.750000 0.750000 B |
| | 0.250000 0.250000 0.750000 B |
| | 0.500000 0.000000 0.750000 B |
| | 0.500000 0.750000 0.750000 B |
| | 0.750000 0.500000 0.750000 B |
| | 0.750000 0.750000 0.500000 B |
| | 0.000000 0.500000 0.000000 B |
| | 0.250000 0.000000 0.500000 B |
| | 0.500000 0.000000 0.000000 B |
| | 0.750000 0.500000 0.000000 B |
| | 0.000000 0.250000 0.500000 C |
| | 0.000000 0.500000 0.500000 C |
| | 0.250000 0.250000 0.250000 C |
| | 0.250000 0.750000 0.500000 C |
| | 0.750000 0.250000 0.000000 C |
| | 0.750000 0.500000 0.250000 C |
| | 0.000000 0.000000 0.250000 C |
| | 0.000000 0.250000 0.000000 C |

Table S3. Atomic volume of BCC Np-U-Zr from DFT and DFT + $U$ at $(U_{\text{eff}}(\text{Np}), U_{\text{eff}}(\text{U}))=(0.6, 0.99)$ eV (unit: eV/atom). The same values are given multiple times where the isopleth paths cross.

| Isopleth path | $x$(Np) | $x$(U) | $x$(Zr) | DFT | DFT + $U$ |
|---|---|---|---|---|---|
| $\text{Np}_x(\text{U}_{0.5}\text{Zr}_{0.5})_{1-x}$ | 0.000 | 0.500 | 0.500 | 21.97 | 22.53 |
|  | 0.250 | 0.375 | 0.375 | 21.74 | 22.52 |
|  | 0.333 | 0.333 | 0.333 | 21.53 | 22.43 |
|  | 0.500 | 0.250 | 0.250 | 20.98 | 22.32 |
|  | 0.750 | 0.125 | 0.125 | 19.53 | 21.74 |
|  | 1.000 | 0.000 | 0.000 | 17.60 | 19.46 |
| $\text{Np}_{0.5}\text{U}_{0.5-x}\text{Zr}_x$ | 0.500 | 0.500 | 0.000 | 19.47 | 20.49 |
|  | 0.500 | 0.375 | 0.125 | 20.13 | 21.56 |
|  | 0.500 | 0.250 | 0.250 | 20.98 | 22.32 |
|  | 0.500 | 0.125 | 0.375 | 21.84 | 22.94 |
|  | 0.500 | 0.000 | 0.500 | 22.49 | 23.26 |
| $\text{Np}_{0.75}\text{U}_{0.25-x}\text{Zr}_x$ | 0.750 | 0.250 | 0.000 | 18.33 | 20.30 |
|  | 0.750 | 0.125 | 0.125 | 19.53 | 21.74 |
|  | 0.750 | 0.000 | 0.250 | 20.90 | 23.14 |
| $\text{U}_x(\text{Np}_{0.5}\text{Zr}_{0.5})_{1-x}$ | 0.500 | 0.000 | 0.500 | 22.49 | 23.26 |
|  | 0.375 | 0.250 | 0.375 | 21.80 | 22.69 |
|  | 0.333 | 0.333 | 0.333 | 21.53 | 22.43 |
|  | 0.250 | 0.500 | 0.250 | 21.06 | 22.04 |
|  | 0.125 | 0.750 | 0.125 | 20.50 | 21.41 |
|  | 0.000 | 1.000 | 0.000 | 20.13 | 20.81 |
| $\text{U}_{0.5}\text{Zr}_{0.5-x}\text{Np}_x$ | 0.000 | 0.500 | 0.500 | 21.97 | 22.53 |
|  | 0.125 | 0.500 | 0.375 | 21.68 | 22.43 |
|  | 0.250 | 0.500 | 0.250 | 21.06 | 22.04 |
|  | 0.375 | 0.500 | 0.125 | 20.28 | 21.51 |
|  | 0.500 | 0.500 | 0.000 | 19.47 | 20.49 |
| $\text{U}_{0.75}\text{Zr}_{0.25-x}\text{Np}_x$ | 0.000 | 0.750 | 0.250 | 21.10 | 21.84 |
|  | 0.125 | 0.750 | 0.125 | 20.50 | 21.41 |
|  | 0.250 | 0.750 | 0.000 | 19.76 | 20.68 |
| $\text{Zr}_x(\text{Np}_{0.5}\text{U}_{0.5})_{1-x}$ | 0.500 | 0.500 | 0.000 | 19.47 | 20.49 |
|  | 0.375 | 0.375 | 0.250 | 21.05 | 22.14 |
|  | 0.333 | 0.333 | 0.333 | 21.53 | 22.43 |
|  | 0.250 | 0.250 | 0.500 | 22.17 | 22.81 |
|  | 0.125 | 0.125 | 0.750 | 22.63 | 22.87 |
|  | 0.000 | 0.000 | 1.000 | 22.92 | 22.92 |
| $\text{Zr}_{0.5}\text{Np}_{0.5-x}\text{U}_x$ | 0.500 | 0.000 | 0.500 | 22.49 | 23.26 |
|  | 0.375 | 0.125 | 0.500 | 22.34 | 23.01 |
|  | 0.250 | 0.250 | 0.500 | 22.17 | 22.81 |
|  | 0.125 | 0.375 | 0.500 | 22.14 | 22.70 |
|  | 0.000 | 0.500 | 0.500 | 21.97 | 22.53 |
| $\text{Zr}_{0.75}\text{Np}_{0.25-x}\text{U}_x$ | 0.250 | 0.000 | 0.750 | 22.80 | 23.09 |
|  | 0.125 | 0.125 | 0.750 | 22.63 | 22.87 |
|  | 0.000 | 0.250 | 0.750 | 22.43 | 22.80 |

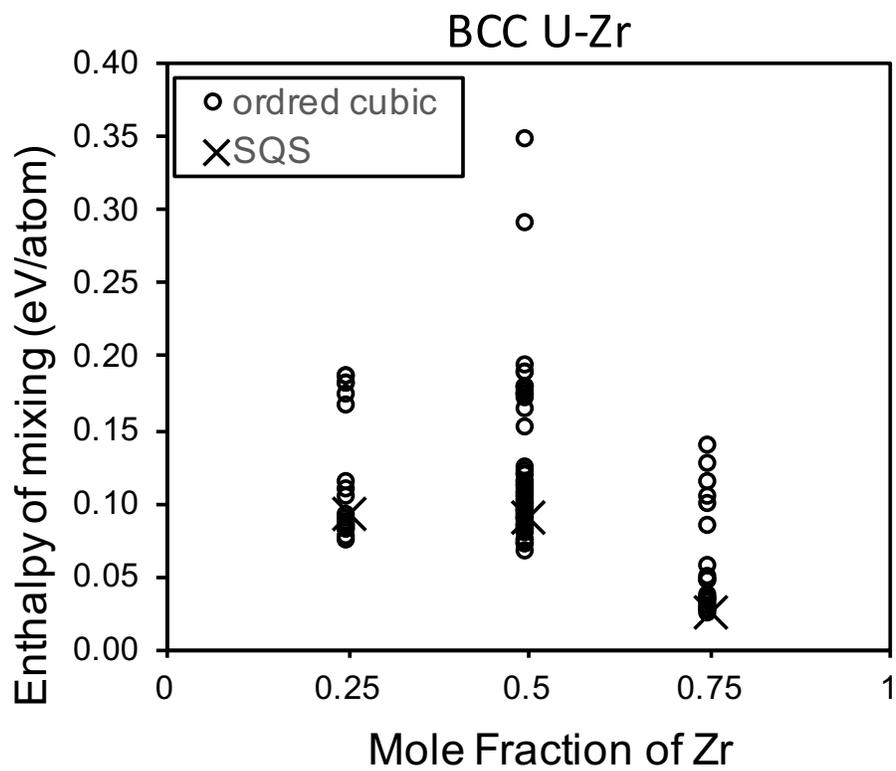

Figure S1. Enthalpy of mixing for BCC U-Zr calculated using the 16-atom SQS supercells used in this study and all symmetrically distinct 16-atom cubic supercells.

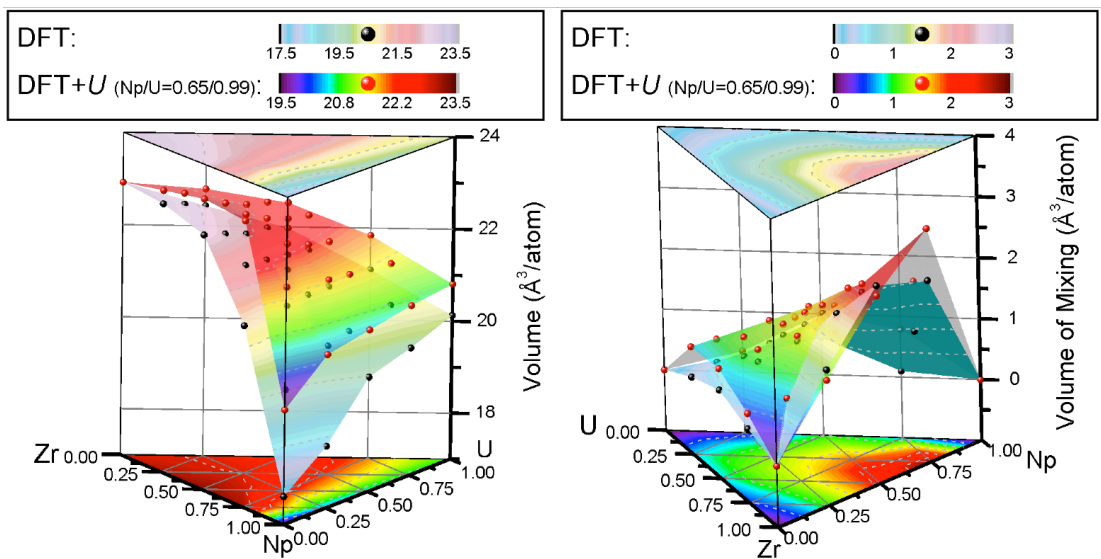

Figure S2. Atomic volume and volume of mixing for BCC Np-U-Zr from DFT and DFT + $U$ at ($U_{\text{eff}}$(Np), $U_{\text{eff}}$(U))=(0.6, 0.99) eV. The front of DFT and DFT + $U$ surface is filled by the color palette given in the legend, while the back of them is filled by dark cyan and gray, respectively. The top and bottom flat surfaces are projections of the DFT and DFT + $U$ 3D surfaces, respectively.

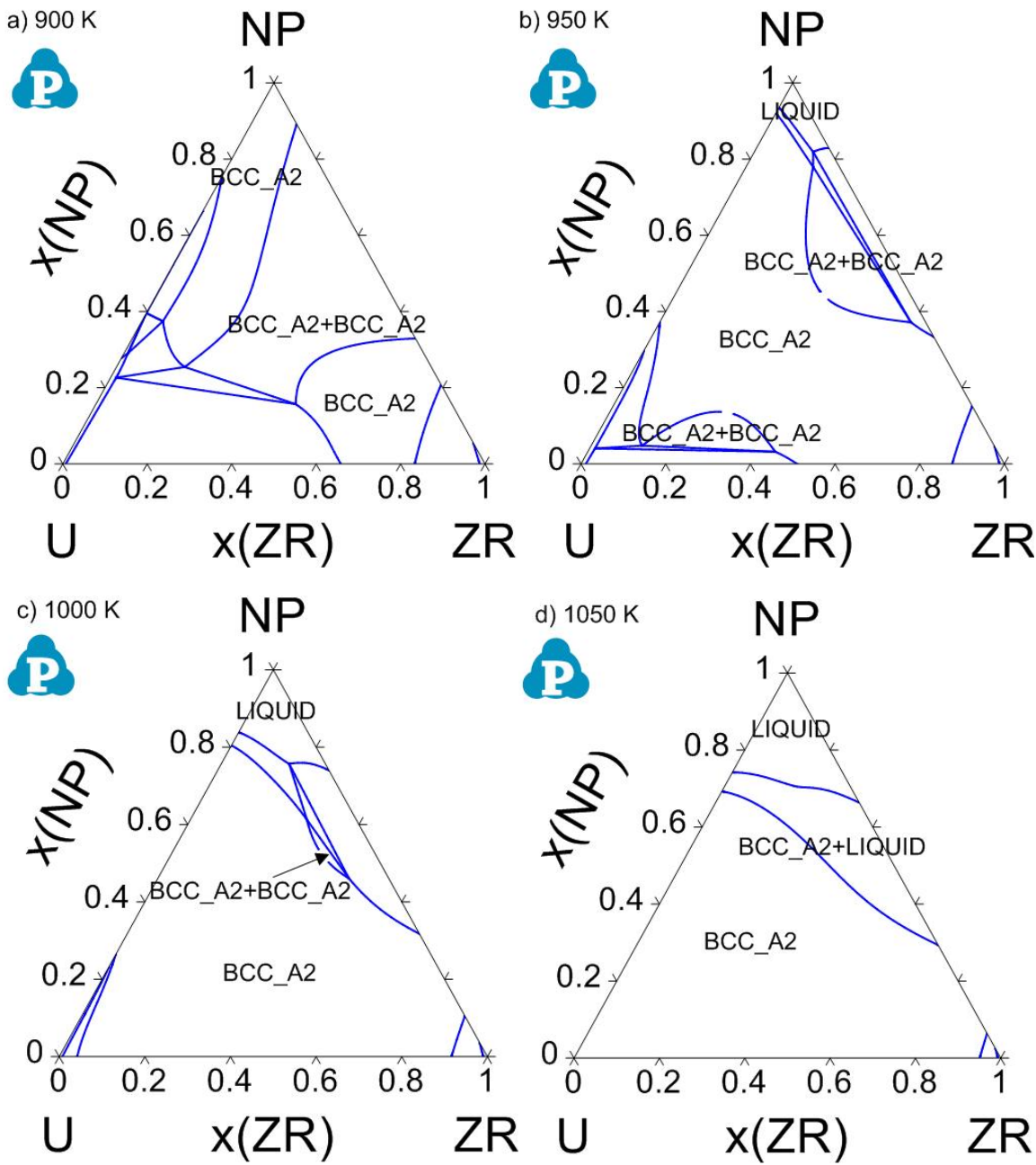

Figure S3. Isothermal sections of Np-U-Zr at a) 900 K, b) 950 K, c) 1000 K and d) 1050 K.